\documentclass[fleqn,usenatbib]{mnras}

\usepackage{newtxtext,newtxmath} 

\usepackage[T1]{fontenc}
\usepackage{ae,aecompl}
\usepackage[caption=false]{subfig}

\usepackage{graphicx}	
\usepackage{amsmath}	
\usepackage{graphicx}	
\usepackage{float,lscape}
\usepackage{comment}
\usepackage[utf8]{inputenc}
\usepackage[dvipsnames]{xcolor}
\usepackage{siunitx}
\usepackage{verbatim}
\usepackage[english]{babel}
\usepackage[T1]{fontenc}
\usepackage{ae,aecompl}
\usepackage{textcomp}
\usepackage[normalem]{ulem}
\usepackage{soul}
\usepackage{lipsum}
\usepackage{booktabs}
\usepackage{url}

\usepackage[inline]{enumitem}

\newcommand*{\email}[1]{\href{mailto:#1}{\nolinkurl{#1}} } 

\numberwithin{equation}{section}

\title{Hunting extreme BL Lacertae blazars with {\it Fermi}-LAT}

\author[M.~Nievas~Rosillo et al.]{\parbox{\textwidth}{
M.~Nievas~Rosillo,$^{1,2,3}\footnotemark[1]$
A.~Dom\'inguez,$^{4}$ 
G.~Chiaro,$^{5,6}$ 
G.~La~Mura,$^{7}$ 
A.~Brill,$^{8}$
and
V.~S.~Paliya$^{9}$ 
}
\vspace{0.4cm}\\  
\parbox{\textwidth}{
$^{1}$ Instituto de Astrof\'isica de Canarias, E-38205 La Laguna, Tenerife, Spain\\
$^{2}$ Universidad de La Laguna, Dept. Astrof\'isica, E-38206 La Laguna, Tenerife, Spain \\
$^{3}$ Deutsches Elektronen-Synchrotron (DESY), Platanenallee 6, Zeuthen, Germany \\
$^{4}$ IPARCOS and Department of EMFTEL, Universidad Complutense de Madrid, E-28040 Madrid, Spain\\
$^{5}$ Institute of Space Astrophysics and Cosmic Physics  IASF / INAF  , Via A.Corti 12, I-20133 Milano Italy \\
$^{6}$ Consorzio Interuniversitario per la Fisica Spaziale  CIFS, Via Pietro Giuria, 1, 10125 Torino IT\\
$^{7}$ Lab. de Instrumenta\c{c}\~ao e F\'{i}sica Experimental de Part\'{i}culas. LIP, Av. Prof. Gama Pinto 2, 1649-003 Lisboa, Portugal \\
$^{8}$ NASA Goddard Space Flight Center, Greenbelt, MD 20771, USA \\
$^{9}$ Aryabhatta Research Institute of Observational Sciences (ARIES), Manora Peak, Nainital 263001, India \\
}}

\date{Accepted XXX. Received YYY; in original form ZZZ}

\pubyear{2021}

\usepackage{newtxtext,newtxmath}

\begin{document}
\newcommand{\de}{\mathrm d}
\label{firstpage}
\pagerange{\pageref{firstpage}--\pageref{lastpage}}
\maketitle

\begin{abstract}
The emission of very-high-energy photons (VHE, $E>100\,\mathrm{GeV}$) in blazars is closely connected to the production of ultra-relativistic particles and the role of these $\gamma$-ray sources as cosmic particle accelerators. This work focuses on a selection of 22 $\gamma$-ray objects from the 2BIGB catalog of high-synchrotron-peaked sources, which are classified as blazar candidates of uncertain type in the 4FGL-DR2 catalog.
We study these sources by means of a re-analysis of the first 10 years of $\gamma$-ray data taken with the {\it Fermi} Large Area Telescope, including the attenuation by the extragalactic background light. Their broadband spectral energy distributions are also evaluated, using multi-wavelength archival data in the radio, optical, and X-ray bands, in terms of one-zone synchrotron-self-Compton models, adding an external Compton component when needed. Out of this analysis, we identify 17 new extreme high-synchrotron-peaked (EHSP) candidates and compare their physical parameters with those of prototypical EHSP blazars. Finally, the resulting models are used to assess their detectability by the present and future generation of ground-based imaging atmospheric Cherenkov telescopes. We find two VHE candidates within the reach of the current and next generation of Cherenkov telescopes: J0847.0-2336 and J1714.0-2029. \\\\

\end{abstract}


\section{Introduction}\label{sec:introduction}

\renewcommand{\thefootnote}{\fnsymbol{footnote}}
\footnotetext[1]{E-mail: \email{mnievas@iac.es}}
\renewcommand{\thefootnote}{\arabic{footnote}}

\begin{figure*}
\label{fig:bcu_location}
\centering
\includegraphics[width=0.9\textwidth]{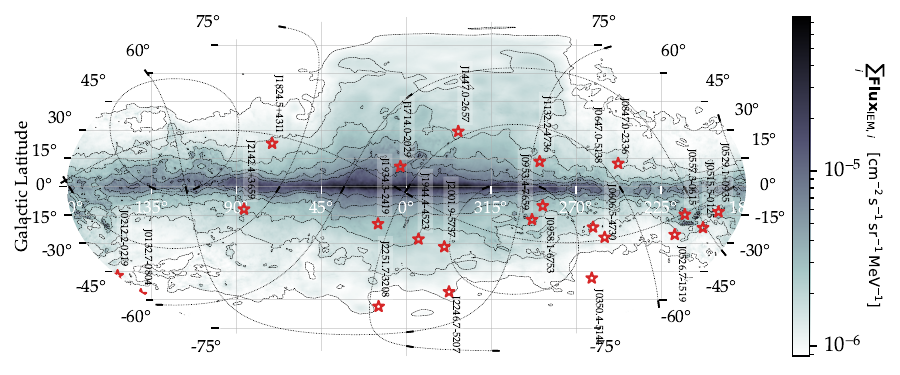}
\caption{Location of the selected sources in Galactic coordinates (red stars). The diffuse emission intensity map from the 4FGL (integrated over energies from $50\,\mathrm{MeV}$ to about $0.8\,\mathrm{TeV}$) is shown as colored contours. Also in dotted black lines we represent the right ascension and declination coordinate system. We analyze only sources at Galactic latitudes of $|b|> 10$~deg in order to limit the effect of the diffuse $\gamma$-ray emission in the modeling of the blazar HE emission and to reduce source confusion issues when looking for lower energy counterparts.}
\end{figure*}

The detection of very-high-energy (VHE, $\mathrm{E}\geq100\,\mathrm{GeV}$) photons has important astrophysical implications since they are a direct proof of extreme cosmic particle acceleration. 
The emission of VHE photons in active galaxies occurs in regions where ultra-relativistic particles are accelerated to energies beyond $10^{18}\,\mathrm{eV}$, 
conditions that are unlikely to be reproduced on Earth. 

Furthermore, the interaction of extragalactic $\gamma$ rays with the optical and infrared (IR) photons from the extragalactic background light (EBL), mainly starlight accumulated since the epoch of re-ionization \citep[e.g.~][]{hauser2001}, is an extremely powerful tool for astronomy. This effect places constraints on galaxy evolution \citep[e.g.~][]{ajello2018_sfh,magic2019_ebl},
cosmology \citep[][]{dominguez2013,biteau2015,dom2019}, and physics beyond the Standard Model \citep[e.g.][]{deangelis2007,sanchezconde2009,abdallah2019,buehler20}.

In the unified model of radio-loud AGN \citep{urry1995}, these sources are basically composed of: 
\begin{enumerate}[leftmargin=*]
\item A central super-massive black hole (SMBH) with mass $\mathrm{M_{SMBH}\sim10^{6-10} M_\odot}$.
\item A sub-pc rotation dominated accretion flow surrounding the SMBH that is usually dubbed {\it accretion disk}. It is formed by material that is falling into the SMBH and can be observed from the optical to X-ray bands with temperatures of $\mathrm{T_{disk}}\sim 10^4\mathrm{K}$.
\item A dusty torus (DT) at $0.1-10\,\mathrm{pc}$ that can induce significant absorption and polarization of the radiation coming out depending on the orientation of the structure with respect to the line of sight. The torus can sometimes be observed in the IR with effective temperatures of $\mathrm{T_{DT}}\sim 10^{2-3}\mathrm{K}$.
\item Gas structures protruding with respect to the accretion plane and giving origin to strong or weak emission lines, which can be broad (BLR) or narrow (NLR) depending on the velocity of the gas.
\item Collimated outflows of energetic particles usually projecting perpendicularly to the disk plane, with tangled magnetic fields and extending up to kpc scales (jets).
\end{enumerate}

It is not well understood how and where the particles are accelerated to ultra-relativistic energies and which specific interactions between these particles and the existing medium produce the observed broadband emission. 
Within the canonical structure of radio-loud active galactic nuclei,   ultra-relativistic particles are typically assumed to be produced either very close to the central object  \citep[e.g.][]{magic_ic310,katsoulakos2018}, in shocks generated from the interaction of the accretion disk with stellar winds or BLR clouds \citep{muller2020} or in knots or substructures embedded in the jets \citep[e.g. ][]{ghisellini2005}. By crossing the shock multiple times, the particles' energy builds up until they escape or energy losses dominate. Alternatively, particle acceleration within magnetic reconnection sites \citep{2020NatCo..11.4176S} and, in some cases, stochastic Fermi acceleration \citep{2017ApJ...842...39L} could be used to explain the origin of ultra-relativistic particles. 

The broadband emission of jet-dominated AGN, and in particular of radio-loud sources with small jet inclinations known as blazars, is usually dominated by non-thermal radiation fields. Blazars have a characteristic double-peaked spectral energy distribution (SED) that extends from radio up to $\gamma$-rays. 
The lower energy component (radio to UV/X-rays) is normally explained as synchrotron emission from ultra-relativistic leptons.
The origin of the high energy component is however still not clear. 
Leptonic models based on inverse Compton scattering of low energy radiation fields \citep{1992ApJ...397L...5M,1994ApJ...421..153S,vanDenBerg2019}, hadronic models with proton-synchrotron or proton-photon interactions leading to the production of secondary decaying mesons, \citep[][and references therein]{1993A&A...269...67M, 2013ApJ...768...54B} or a combination of both are possible scenarios. Yet neither of these scenarios is able to successfully predict other observables, for example neutrinos in leptonic models and fast (minute) variability in hadronic models. 

The host emission, if visible, mostly contains starlight and possibly re-emission from dust. The radiation is found typically in the optical/IR range, building up over the synchrotron spectrum from the relativistic jets. 
In addition to the non-thermal emission, some of these AGN have thermal components that can locally outshine the non-thermal spectrum anywhere in the IR to ultraviolet band. Examples of these components are the IR emission from the DT, the optical/UV emission from the accretion disk, the radiation from the X-ray corona at either the base of the jet or the inner accretion flow, and the emission lines arising from the NLR and the BLR.
The intensity of such features, particularly the rest-frame width of the optical emission lines, divides blazars into BL Lacertae (BLL) objects and flat spectrum radio quasars \citep[FSRQs, see ][]{urry1995,2011MNRAS.414.2674G}. BLL have a continuum dominated optical spectrum. Emission lines, if present at all, are typically weak. FSRQs on the other hand often exhibit strong and broad (equivalent width $\mathrm{EW} > 5\,\mathring{A}$) emission lines in the optical regime \citep[see, e.g.,][]{2011MNRAS.414.2674G,paliya21}.

Blazars can also be sub-classified according to the position of the frequency of the peak of their synchrotron emission spectrum \citep{abdo10} in:  
\begin{enumerate*}[label=(\roman*),labelwidth=0pt] 
\item low-synchrotron-peaked (LSP, $\mathrm{\nu_{SP}}<10^{14}\,\mathrm{Hz}$), 
\item intermediate-synchrotron-peaked (ISP, $10^{14}\,\mathrm{Hz} \leq\mathrm{\nu_{SP}}<10^{15}\,\mathrm{Hz}$, 
\item high-synchrotron-peaked (HSP, $10^{15}\,\mathrm{Hz} \leq\mathrm{\nu_{SP}}<10^{17}\,\mathrm{Hz}$) and 
\item extreme high-synchrotron-peaked (EHSP, $\mathrm{\nu_{SP}}\geq10^{17}\,\mathrm{Hz}$). 
\end{enumerate*}
The LSP and ISP groups contain both BLLs and FSRQs, whereas HSPs are predominantly BLLs. 
As a result, there is a connection between the spectral classification and the presence of features in the optical spectrum, i.e., HSPs have predominantly featureless spectra, as opposed to LSPs and ISPs. 
Moreover, HSPs tend to be less variable than LSPs and ISPs in the high-energy (HE, $100\,\mathrm{MeV}\leq\mathrm{E}<100\,\mathrm{GeV}$) $\gamma$-ray band. 
However, this result could be an observational bias since they are usually fainter in that band. 
The smooth transition in the properties from HSPs to FSRQs is known as the blazar sequence \citep{padovani2007,ghisellini2017}, 
but we note that observational biases \citep[e.g.~][]{giommi2012} are substantial and the blazar sequence may not be valid \citep{keenan2021,1wshp,3hsp}. 

EHSPs are often regarded as promising VHE emitters and searching for new sources of this class is of particular importance for TeV instruments \citep{2020ApJS..247...16A,2021ApJ...916...93Z}. However, their detection with current $\gamma$-ray instruments is challenging. On one hand, survey-mode instruments such as {\em Fermi}-LAT are often not sensitive enough at the energies where EHSP have the maximum of their emission. In addition, their low luminosity and the lack of strong variability makes it even harder for {\em Fermi}-LAT to detect many of these sources. As a result, very few EHSPs are known as $\gamma$-ray emitters and long exposures are often required to detect them with both {\em Fermi}-LAT and imaging atmospheric Cherenkov telescopes (IACTs). Because IACTs have low duty cycles, the selection of the most promising targets becomes crucial.
Yet understanding these sources is important for diverse science topics, such as the study of the EBL \citep[e.g.~][]{dominguez15} and the diffuse cosmic $\gamma$-ray background \citep[e.g.~][]{paliya19}. 

In this work, we use the existing 4LAC-DR2 and 2BIGB $\gamma$-ray catalogs to identify extreme blazar candidates from known blazars of unknown source class. 
Second, we re-analyze {\em Fermi}-LAT data from each source position in order to include the attenuation by the EBL. 
Third, we derive their physical properties based on a broadband SED modeling, and we provide a new spectral classification for the sources. 
Finally, we discuss their possible emission in the TeV band based on direct extrapolations for the SED modeling within a multi-wavelength (MWL) context in addition to the $\gamma$-ray data. From this analysis, we propose two EHSP blazar candidates which may be detectable at TeV energies: J0847.0-2336 and J1714.0-2029.

This document is structured as follows: Source selection is presented in section \ref{sec:sample}, including the search for redshift measurements and low energy counterparts for each source and the $\gamma$-ray data analysis. The modeling of the multiwavelength-emission is described in section \ref{sec:method}, introducing the theoretical framework of one-zone leptonic models and presenting the best-fit parameters. The detectability prospects of the blazar sample with existing and future IACTs is shown in section \ref{sec:vhe_prospects}. Finally, we discuss the main results of this work in section \ref{sec:discussion}. Throughout this document, we assumed a flat $\Lambda$CDM cosmology with a Planck constant $H_0=67.8\,\mathrm{km\,s^{-1} Mpc^{-1}}$, matter density parameter $\Omega_{m,0}=0.307$, baryon density parameter $\Omega_{b,0} = 0.0483$, a thermal black body temperature of the CMB $T_{CMB,0} = 2.725\,\mathrm{K}$, effective number of relativistic degrees of freedom $N_{\rm eff} = 3.05$ and sum of neutrino masses of $0.06\,\mathrm{eV}$.

\section{Source selection and Data analysis}\label{sec:sample}
\subsection{Source selection}

The Large Area Telescope (LAT), onboard the {\em Fermi Gamma-ray Space Telescope}, is an imaging, wide field-of-view (FoV) instrument that uses the pair-production technique to detect $\gamma$-rays in the energy range from below 20 MeV to above 300 GeV \citep{fermilat}. 
The 4FGL catalog contains more than 3200 AGN detected by LAT in its first eight years of operation \citep{4fgl}. The observations span from 2008 Aug 4 to 2016 Aug 2, and cover the energy range from $50\,\mathrm{MeV}$ to $1\,\mathrm{TeV}$. Based on the same data used to build the 4FGL, a dedicated AGN catalog named 4LAC was released, listing additional properties such as redshift estimates \citep{4lac}. 
Both the 4FGL and the 4LAC were subsequently updated with a ``Data Release 2'', which extends the total telescope time to 10 years \citep{ballet20}.

This updated release adds new $\gamma$-ray objects, improving also the identification and classification of already existing ones, supported with data from more recent surveys and targeted observations across multiple bands.
With a detection threshold set at a Test Statistic (TS) of 25, the current number of sources listed in the 4LAC-DR2 as BLLs and FSRQs is 1308 and 744, respectively. Furthermore, 1384 sources remain classified as blazar candidates of uncertain type (BCU). These are objects with broadband characteristic of blazars but lacking a clear optical spectroscopic confirmation.

This work begins with the identification of the most promising HSP emitters among BCUs from the {\em Fermi}-LAT 4LAC-DR2 catalog with the aim of determining their spectral class following a multi-wavelength modeling approach. To do so, we cross-match the entire BCU list from 4LAC-DR2 with the 2BIGB catalog \citep{2bigb}. 2BIGB, built on top of another catalog named 3HSP \citep{3hsp}, and other previous releases \citep{2017A&A...598A.134A, 2018MNRAS.480.2165A}. The 3HSP catalog includes a new analysis of $\gamma$-ray data from LAT, taken over the first 11 years of mission, at locations of infrared sources with similar properties to HSP objects. 2BIGB lists 1160 sources, with $\mathrm{\nu_{SP}}>10^{15}\,\mathrm{Hz}$, that have been clearly detected in HE $\gamma$-rays by \cite{2bigb}. As a result, 2BIGB provides a cleaner list of sources to perform our search than if directly using the 2013 sources from 3HSP. Since 3HSP already uses the absence of emission lines in the optical band as a selection criterion, both 3HSP and 2BIGB should be dominated by BL Lac sources.

To select the most suitable candidates for present and future VHE observations, we incorporated a ``Figure of Merit'' ($\mathrm{FOM}$) cut of $\mathrm{FOM} > 0.7$. $\mathrm{FOM}$ was originally defined in the 1WHSP catalog \citep{1wshp} as the ratio between the flux of the synchrotron peak of the considered source in the SED and the flux at the synchrotron peak of the faintest blazar detected at the time in TeV energies (4FGL~J0013.9-1854). This definition was subsequently updated in the 3HSP catalog as the number of VHE sources increased. 3HSP's definition of FOM, used throughout this document, is referred to 4FGL~J0014.1-5022, whose energy flux at the synchrotron peak is $2.5 \times 10^{-12}\, \mathrm{erg\, s^{-1}\,cm^{-2}}$. 

Finally, to allow for a proper modeling of the physics in the jets, we keep only sources with good multi-wavelength coverage using the {\em sflag} from 2BIGB, and also measured synchrotron peak $\mathrm{\nu_{SP}}$ or at least a lower limit ({\tt nuflag $\in$ [1,3])}. We also require at least a redshift estimate, being it either spectroscopic or photometric, as described in section \ref{sec:redshift}. Sources within $\pm 10\,\mathrm{deg}$ in Galactic latitude (i.e., near the Galactic plane) are not considered in the study as their classification is likely more problematic: source confusion and bright $\gamma$-ray diffuse components can bias the spectral reconstruction of the source, particularly for those with the lowest $\gamma$-ray fluxes. All these requirements considered, we end up selecting 22 sources for our ``Master Sample'' (plus J0733.4$+$5152, detected in 2018 by MAGIC \citep{j0733} and therefore excluded from further analysis), listed in Table \ref{tab:master_sample} and shown on a sky-map in Figure~\ref{fig:bcu_location} as red stars over the contours representing the integral diffuse $\gamma$-ray emission seen by {\em Fermi}-LAT.

\subsection{Redshift determination}\label{sec:redshift}

Redshifts are extracted from 4LAC \citep{4lac}, and cross-checked and updated using the results from \cite{3hsp} and \cite{2020arXiv201205176G}. For sources lacking spectroscopic redshift measurements, we used photometric estimates from \cite{3hsp} ({\em zflag}=5). The latter were obtained by fitting a magnitude $M_R=-23.5$ giant elliptical galaxy template to the optical photometric data.

\begin{table}
\caption{Master sample as obtained from 2BIGB. The positional information, in equatorial coordinates, corresponds to the $\gamma$-ray sources as they appear in the 4LAC-DR2. 
Redshifts ($z$) were extracted from \citet{4lac,3hsp,2020arXiv201205176G}.
When a photometric redshift was used, we appended the flag `h' to the redshift value, standing for `host-fitting photometric redshift'. 
TS refers to the test statistics as presented in the 2BIGB analysis. 
Finally, FOM is the Figure of Merit as is portrayed in the 3HSP and the 2BIGB catalogs. 
All these sources are classified as BCUs in the 4FGL-DR2. 
$\dagger$: J0733.4$+$5152 was detected in 2018 by MAGIC \protect\citep{j0733} and therefore is excluded from further analysis.}
\label{tab:master_sample}
\begin{tabular}{lrrrrr}
\toprule
    4FGL Name &   RAJ2000 &   DEJ2000 &   $z$ &     TS &  FOM \\
\midrule
 J0132.7$-$0804 &    23.183 &    -8.074 &  0.148  &     88 &  0.8 \\
 J0212.2$-$0219 &    33.066 &    -2.319 &  0.250  &     61 &  0.8 \\
 J0350.4$-$5144 &    57.613 &   -51.743 &  0.32h &     98 &  0.8 \\
 J0515.5$-$0125 &    78.891 &    -1.419 &  0.25h &     55 &  0.8 \\
 J0526.7$-$1519 &    81.692 &   -15.321 &  0.21h &    218 &  1.6 \\
 J0529.1$+$0935 &    82.297 &     9.597 &  0.30h &     86 &  1.3 \\
 J0557.3$-$0615 &    89.344 &    -6.265 &  0.29h &     53 &  1.6 \\
 J0606.5$-$4730 &    91.642 &   -47.504 &  0.030  &    137 &  1.0 \\
 J0647.0$-$5138 &   101.773 &   -51.638 &  0.22h &     81 &  2.5 \\
 J0733.4$+$5152$^\dagger$ &   113.362 &    51.880 &  0.065  &    162 &  2.5 \\
 J0847.0$-$2336 &   131.757 &   -23.614 &  0.059  &    921 &  0.8 \\
 J0953.4$-$7659 &   148.367 &   -76.993 &  0.25h &    104 &  0.8 \\
 J0958.1$-$6753 &   149.534 &   -67.894 &  0.21h &     29 &  1.0 \\
 J1132.2$-$4736 &   173.056 &   -47.613 &  0.21h &    129 &  1.0 \\
 J1447.0$-$2657 &   221.765 &   -26.962 &  0.32h &     46 &  2.0 \\
 J1714.0$-$2029 &   258.522 &   -20.486 &  0.09h &    110 &  2.0 \\
 J1824.5$+$4311 &   276.126 &    43.196 &  0.487  &     99 &  0.8 \\
 J1934.3$-$2419 &   293.582 &   -24.326 &  0.23h &     63 &  1.6 \\
 J1944.4$-$4523 &   296.101 &   -45.393 &  0.21h &    164 &  1.0 \\
 J2001.9$-$5737 &   300.491 &   -57.631 &  0.26h &    123 &  0.8 \\
 J2142.4$+$3659 &   325.602 &    36.986 &  0.24h &    110 &  1.3 \\
 J2246.7$-$5207 &   341.682 &   -52.126 &  0.098  &     95 &  2.5 \\
 J2251.7$-$3208 &   342.944 &   -32.140 &  0.246  &     52 &  2.0 \\
\bottomrule
\end{tabular}
\end{table}

\begin{figure}
\centering
\includegraphics[width=0.45\textwidth]{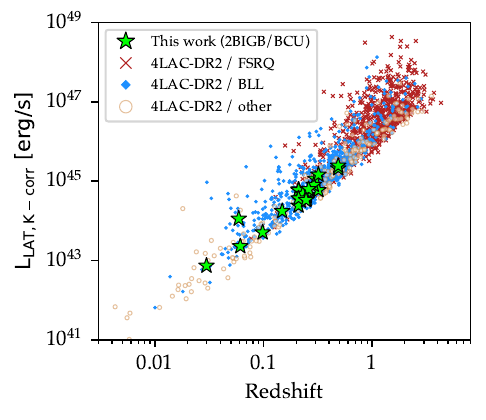}
\caption{
Luminosity distribution of blazars as a function of redshift for BL Lac objects (blue diamonds) and FSRQs (red crosses). In green stars the BCU that we selected for this study. For completeness, other AGN (including BCUs) with redshift information in the 4LAC-DR2 are shown as open brown circles.
}\label{fig:blazar_sequence}
\end{figure}

Table \ref{tab:master_sample} includes the redshift estimates for the 22 selected targets (23 including J0733.4$+$5152). Figure \ref{fig:blazar_sequence} shows the K-corrected luminosity versus redshift, using the fluxes and spectral indices included in the 4LAC and the approximate K-correction $(1+z)^{\alpha-2}$ following \cite{2012ApJ...751..108A}. As can be seen, our sources have redshifts and luminosities consistent with those of typical BLL.

\begin{table}
\caption{Photon spectral index, variability index and fractional variability of the master sample, extracted from the 4LAC-DR2. $\dagger$: J0733.4$+$5152 was detected in 2018 by MAGIC \protect\citep{j0733} and therefore is excluded from further analysis.}
\label{tab:variability}
\begin{tabular}{lrrrr}
\toprule
4FGL Name &  Spectral index &  Var. index &  Frac. variability \\
\midrule
 J0132.7$-$0804  & $1.82 \pm 0.11$ &  1.23 & - \\
 J0212.2$-$0219  & $2.14 \pm 0.15$ & 15.30 & $0.50 \pm 0.31$ \\
 J0350.4$-$5144  & $1.85 \pm 0.13$ &  7.45 & $0.18 \pm 0.73$ \\
 J0515.5$-$0125  & $1.96 \pm 0.14$ & 11.53 & $0.34 \pm 0.33$ \\
 J0526.7$-$1519  & $1.96 \pm 0.08$ &  7.48 & $0.11 \pm 0.37$ \\
 J0529.1$+$0935  & $1.98 \pm 0.13$ & 12.53 & $0.39 \pm 0.27$ \\
 J0557.3$-$0615  & $1.87 \pm 0.15$ &  6.12 & - \\
 J0606.5$-$4730  & $2.01 \pm 0.10$ &  9.55 & $0.21 \pm 0.25$ \\
 J0647.0$-$5138  & $1.83 \pm 0.14$ & 11.02 & $0.20 \pm 0.60$ \\
 J0733.4$+$5152$^\dagger$  & $1.80 \pm 0.10$ & 14.97 & $0.48 \pm 0.25$ \\
 J0847.0$-$2336  & $1.94 \pm 0.04$ & 12.37 & $0.15 \pm 0.10$ \\
 J0953.4$-$7659  & $1.91 \pm 0.16$ &  5.33 & - \\
 J0958.1$-$6753  & $2.04 \pm 0.20$ & 11.50 & $0.49 \pm 0.64$ \\
 J1132.2$-$4736  & $2.03 \pm 0.09$ &  7.30 & $0.09 \pm 0.54$ \\
 J1447.0$-$2657  & $1.87 \pm 0.15$ &  7.36 & - \\
 J1714.0$-$2029  & $1.63 \pm 0.12$ & 22.72 & $0.69 \pm 0.31$ \\
 J1824.5$+$4311  & $1.92 \pm 0.14$ &  6.26 & - \\
 J1934.3$-$2419  & $1.84 \pm 0.12$ &  8.50 & - \\
 J1944.4$-$4523  & $1.70 \pm 0.11$ & 10.69 & $0.16 \pm 0.54$ \\
 J2001.9$-$5737  & $2.10 \pm 0.11$ &  5.00 & - \\
 J2142.4$+$3659  & $1.97 \pm 0.13$ & 10.79 & $0.37 \pm 0.33$ \\
 J2246.7$-$5207  & $1.61 \pm 0.13$ & 15.59 & $0.55 \pm 0.34$ \\
 J2251.7$-$3208  & $1.72 \pm 0.16$ &  7.07 & - \\
\bottomrule
\end{tabular}
\end{table}

\subsection{Search for archival radio, optical and X-ray data}\label{sec:mw-archival}

To better constrain the source types and put their broadband emission in context, we collected archival observations for the selected sources using the Space Science Data Center (SSDC) {\tt SED Builder} service\footnote{\protect\url{https://tools.ssdc.asi.it/SED/}}, which contains multi-wavelength data from several instruments and surveys acquired over decades. We set the maximum search radius to $5\,\mathrm{arcmin}$ centred at the position of the low-energy counterpart that is reported for each $\gamma$-ray point source in the 4LAC-DR2 catalog. 
We included in the search all the data, not covered by {\em Fermi}-LAT, that was available in SSDC for each source, without time constraints (e.g., we did not exclude periods with possible flaring activity). For some $\gamma$-ray sources, the search circle resulted in more than one possible low-energy counterpart. In such cases, we selected the candidate counterpart that had the best instrument coverage and performed a sanity check consisting of a visual inspection of the shape of the SED to look for features typical of (E)HSP blazars. These features may be increasing flux from radio to UV/X-rays, typical double-peaked spectral shape from blazars, and presence of an excess in the optical compatible with that of a host galaxy. 
Some of the instruments and surveys covered in this search are 1SWXRT \citep{1SWXRT}, 1SXPS \citep{1SXPS}, 2MASS \citep{2MASS}, ARGO2LAC \citep{ARGO2LAC}, AT20GCAT \citep{AT20GCAT}, ATCAPMN \citep{ATCAPMN}, ATPMNCAT \citep{ATPMNCAT}, BAT60AGN \citep{BAT60AGN}, BATPA100 \citep{BATPA100}, Catalina RTS \citep{CRTS}, DENIS\_3 \citep{DENIS3}, FIRST \citep{FIRST}, GALEXAISFUV \citep{GALEX}, GB6 \citep{GB6}, IPCSLEW \citep{IPCS}, NED,\protect\footnote{The NASA/IPAC Extragalactic Database (NED) is operated by the Jet Propulsion Laboratory, California Institute of Technology, under contract with the National Aeronautics and Space Administration.} NVSS \citep{NVSS}, PMN \citep{PMN2,PMN3,PMN1,PMN4}, RXS2CAT \citep{RXS2CAT}, SDSS13 \citep{SDSS13}, SDSS7 \citep{SDSS7}, SUMSS \citep{SUMSS}, SWBAT105 \citep{SWBAT105}, USNO A2.0 \citep{USNOA2}, UVOTSSC \citep{UVOTSSC}, VLSSr \citep{VLSSr}, VizieR photometry \citep{VIZIER}, WGACAT2 \citep{WGACAT}, WISE \citep{ALLWISE} and XMMSL1D6 \citep{XMMSL1}.

\subsection{Gamma-ray data analysis}\label{sec:obs_analysis}

\subsubsection{Event selection}

For each target, we considered only Pass 8 \citep{atwood2013,bruel2018} source-class events detected in a region of interest (ROI) of $20^\circ$ radius centred on the nominal position of the target, which we take from 4FGL-DR2.
We use events extending from $100\,\mathrm{MeV}$ to $1.5\,\mathrm{TeV}$.  
The upper energy limit is about the highest energy we can cover with the published Pass 8 instrument response functions (IRFs) if we take into account energy dispersion corrections. Together with the inclusion of EBL absorption in the source modeling, this allows to extrapolate the spectrum to VHE with greater accuracy than if EBL absorption is not taken into account and energy dispersion corrections are ignored.
We followed the event selection recommendations from the {\it Fermi}-LAT analysis {\em Cicerone}\footnote{\protect\url{https://fermi.gsfc.nasa.gov/ssc/data/analysis/documentation/Cicerone/}}, including only good quality data {\tt [(DATA\_QUAL>0)\&\&(LAT\_CONFIG=1)]}. To be consistent with the 4FGL-DR2, we used the same Good Time Intervals (GTI) and temporal coverage that were considered in that catalog, that is August 4, 2008 to August 2, 2018. The selection cuts and data analysis strategy are similar to 4FGL-DR2. The differences are the inclusion of events from $100\,\mathrm{MeV}$ (instead of $50\,\mathrm{MeV}$ as in 4LAC-DR2) up to higher energies of $1.5\,\mathrm{TeV}$, the addition of EBL attenuation in the spectral model of the source of interest, 
and the construction of the sky model for the rest of the sources in the field: instead of procedurally generating the sky model by creating seed sources to reproduce excess count clusters in the skymap as in the 4FGL-DR2, we use the 4FGL-DR2 catalog to fetch the position of the sources (either point-like and extended) and their spectral parameters. We then only free the spectral parameters of those sources that have high detection significance or are located near the source of interest. 
The analysis of the data considered the average emission from the sources over the ten years of telescope time, without discarding any particular time interval during which the source may have undergone a flare. The data of each sample were reduced and analyzed using the open-source software package {\tt enrico} \citep{enrico2013}, a wrapper written in python that uses internally the official {\tt Fermitools} \citep[][ version 2.0.0]{fermitools} installed through {\tt conda} \citep{anaconda}. 

\subsubsection{Binned likelihood analysis}

We applied a binned likelihood analysis approach using point spread function (PSF) event types following the same strategy as in the 4FGL-DR2, i.e. considering only the photons with the best angular resolution in the lowest energies and gradually adding lower quality event types as energy increases. Each event type was analyzed using its own set of IRFs and then combined at the likelihood maximization stage.
We used 10 bins per energy decade and IRFs {\tt P8R3\_SOURCE\_V3\_v1}. The likelihood model was built including all point-like and extended sources available in the 4FGL-DR2 within the selected field of view plus an additional $10^\circ$ ring concentric to that field. This `buffer' is needed to account for bright sources outside our FoV which may have strong contaminating tails entering the ROI. The spectral parameters and morphology were fixed to those of the 4FGL for sources outside a radius of $8^\circ$. The exception are those with a high TS ($\mathrm{TS}>25$), for which we free the normalization of the spectrum. Sources within the inner $8^\circ$ had their spectral parameters free to cover any possible flare or spectral change that could contaminate the measurement of the flux for the source of interest.

\subsubsection{Spectral models}

The non-thermal nature of the emission in the higher energy component of the blazar broadband SED makes it possible to represent its spectrum with simple analytical functions, for instance a power-law (PWL), also possibly with an exponential cut-off (EPWL) or a log-parabola (LP). 
For sources found at cosmological distances ($z>0$), the $\gamma$-ray radiation is absorbed in the interaction with EBL photons \citep{gould67,1992ApJ...390L..49S}, with a characteristic optical depth ($\tau_{\rm EBL}$) that depends on the distance and energy of the incident $\gamma$-ray \citep[e.g.~][]{dom,saldana-lopez21}. The generalized form of these spectral shapes is:

\begin{eqnarray}
\frac{\de N(E)}{\de E} = N_0 \left( \frac{E}{E_0} \right)^{-[\alpha + \beta \mathrm{log}\,(E/E_0)]} \mathrm{e}^{-[\tau_{\rm EBL}(z,E) + (E / E_{\rm cut})^{\Gamma}]} \label{eqSEDform}
\end{eqnarray}

\noindent
where $\alpha$ is the photon or spectral index, which is often assumed to be $\alpha \ge 1.5$ for both shock-accelerated electron acceleration plus inverse Compton emission and for proton-plasma interactions \citep{2001RPPh...64..429M,2006Natur.440.1018A}. 
$\beta$ is the curvature parameter in the case of LP models. $E_0$ is the pivot energy, which in our case is fixed at the decorrelation energy, or energy at which the flux errors reach their minimum value, for each source independently. $E_{\rm cut}$ is the cut-off energy for models with cut-offs, e.g. EPWL. Finally $\Gamma$ is the index or strength of the cut-off. Setting $\beta=0$ or $E_{\rm cut}=\infty$ allows the recovery of an EPWL or LP respectively, while setting both values simultaneously to $\beta=0$ and $E_{\rm cut}=\infty$ provides the canonical PWL. As a reference, Table \ref{tab:variability} provides the spectral indices as they appear in the 4LAC-DR2  for the PWL case. 

For each of our candidates, we repeat the data analysis assuming three possible analytical shapes to model the spectrum: a simple PWL, a LP, and a EPWL (with $\Gamma=1$), all of them absorbed by EBL. The reason is that we are interested in estimating the TeV detectability of our sources, and these models lead to different predictions. In general, we find that the extrapolation towards very high energies of a PWL leads to overly optimistic estimation of VHE fluxes and consequently we ended up removing it from the pool of allowed spectral shapes. Conversely, the EPWL results in the lowest predicted VHE fluxes, as expected because of the strong downward curvature. We note that the 4FGL (and 4LAC) catalog considers instead the PLEC, defined as a Power Law with an exponential cut-off index of $\Gamma=2/3$. While possibly a better match for typical blazar spectra, we wanted to explore a more conservative model, with a cut-off index of $\Gamma=1$, in marked contrast with the LP case. The resulting {\em Fermi}-LAT data analyses using the two models described, for each source, are presented in Figure \ref{fig:lat_seds}. As a reference, the spectral index registered in the 4LAC is reported in Table \ref{tab:variability}.

\begin{figure*}
\centering
\includegraphics[width=0.325\linewidth]{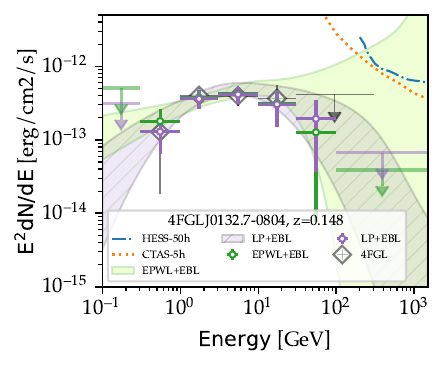}
\includegraphics[width=0.325\linewidth]{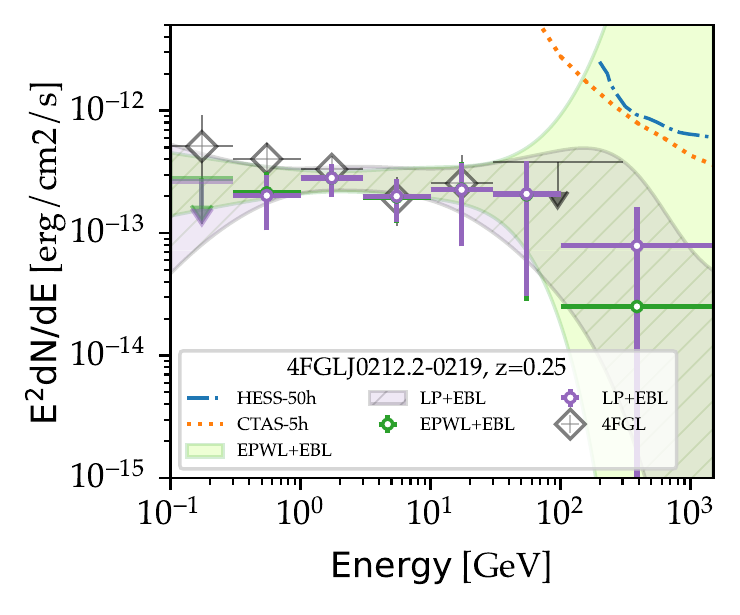}
\includegraphics[width=0.325\linewidth]{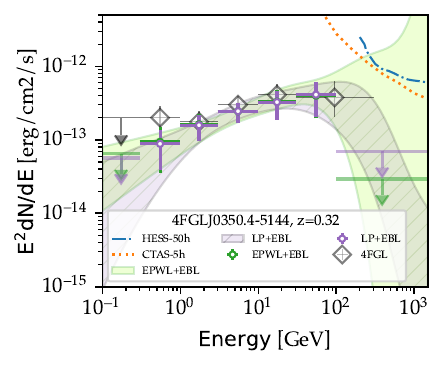}
\includegraphics[width=0.325\linewidth]{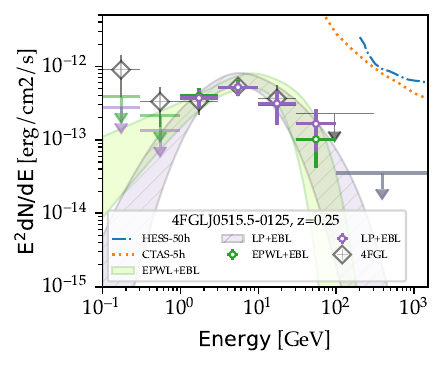}
\includegraphics[width=0.325\linewidth]{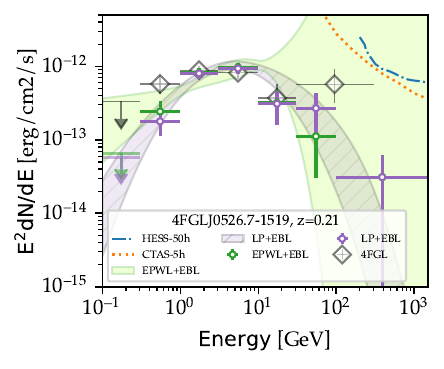}
\includegraphics[width=0.325\linewidth]{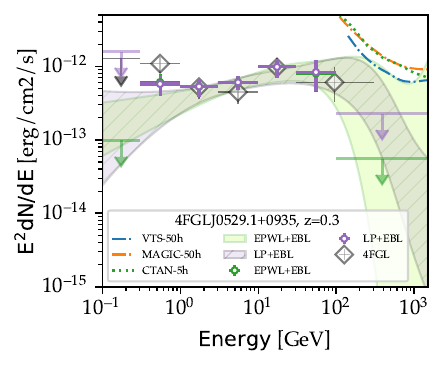}
\includegraphics[width=0.325\linewidth]{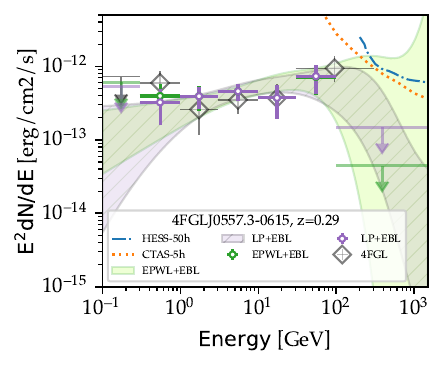}
\includegraphics[width=0.325\linewidth]{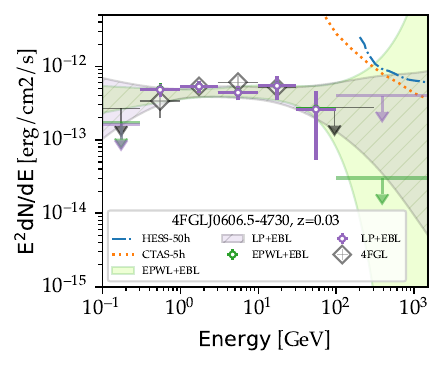}
\includegraphics[width=0.325\linewidth]{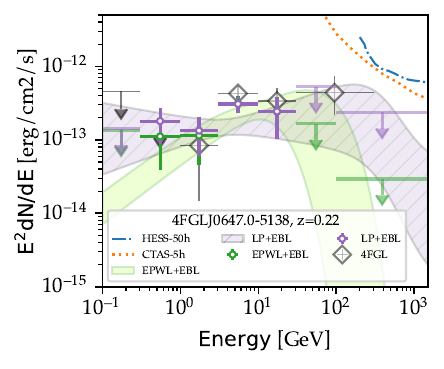}
\includegraphics[width=0.325\linewidth]{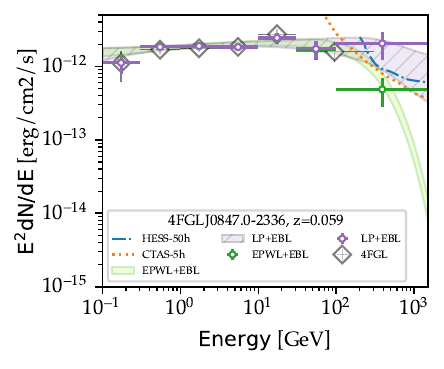}
\includegraphics[width=0.325\linewidth]{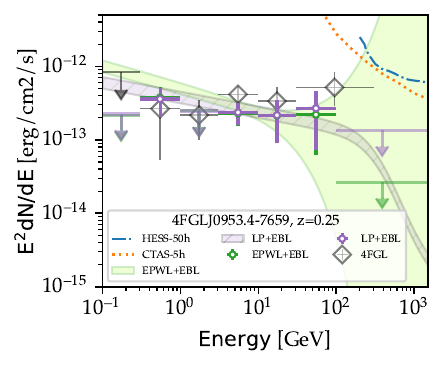}
\includegraphics[width=0.325\linewidth]{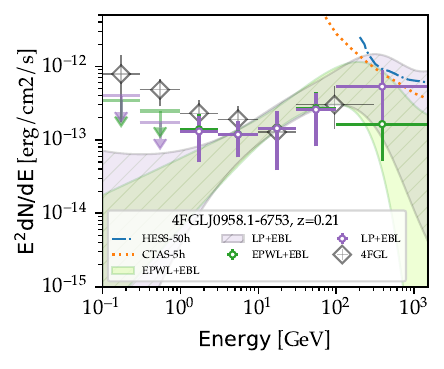}
\includegraphics[width=0.325\linewidth]{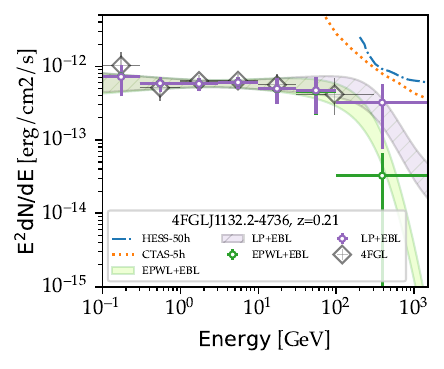}
\includegraphics[width=0.325\linewidth]{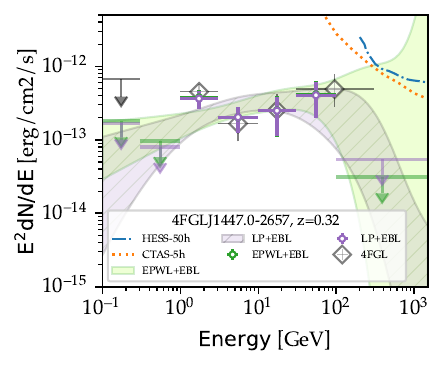}
\includegraphics[width=0.325\linewidth]{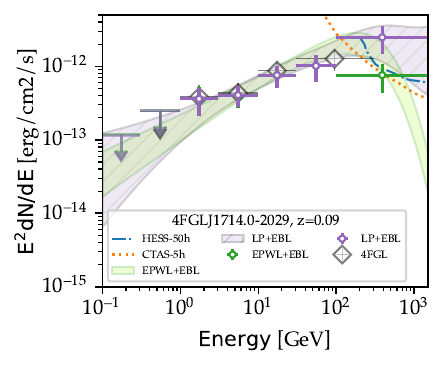} 
\label{fig:lat_seds0}
\end{figure*}

\begin{figure*}
\ContinuedFloat
\centering
\includegraphics[width=0.325\linewidth]{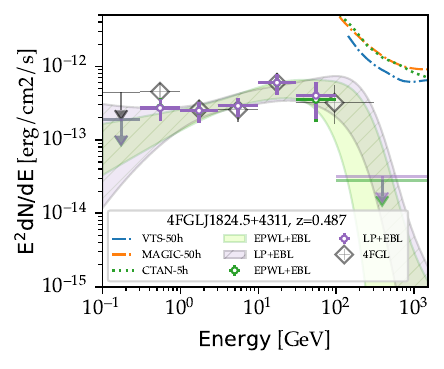}
\includegraphics[width=0.325\linewidth]{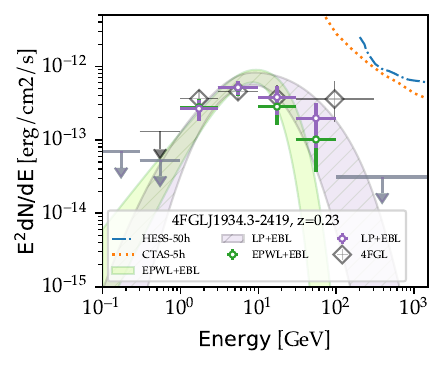}
\includegraphics[width=0.325\linewidth]{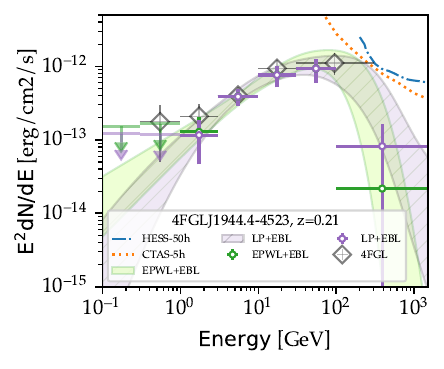} 
\includegraphics[width=0.325\linewidth]{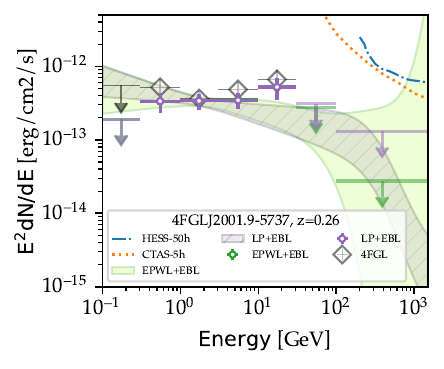}
\includegraphics[width=0.325\linewidth]{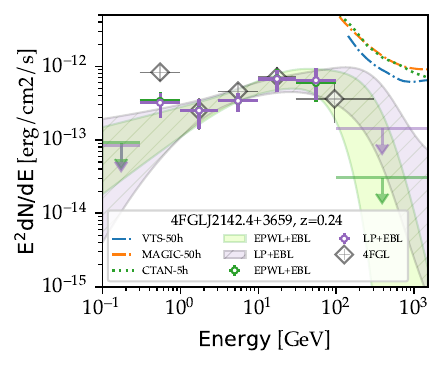}
\includegraphics[width=0.325\linewidth]{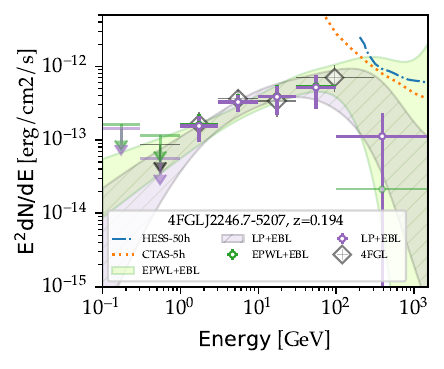} 
\includegraphics[width=0.325\linewidth]{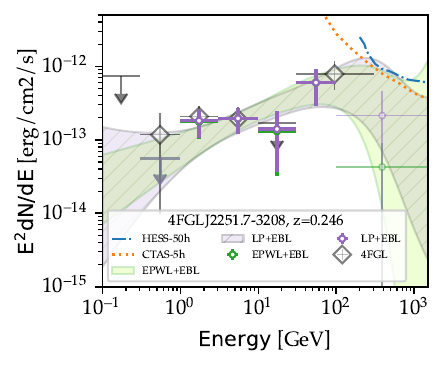}
\caption{Spectral energy distribution of the selected HSP BCUs from the 2BIGB, including the EBL absorption effects for the assumed blazar redshifts of Table \ref{tab:master_sample}. The bow-tie plots for two curved models (Log Parabola or LP, in purple dashed, and Power Law with exponential cut-off or EPWL in solid light green) are shown together with the spectral points up to an energy of $1.5\,\mathrm{TeV}$, presented in purple and green. For consistency, the original analysis presented in the 4FGL catalog is presented as gray points. Upper limits are calculated and reported when the significance of the energy bin is lower than $1\,\sigma$. Bins with very low significance, whose negative errors are larger than $90\%$ of the value, as calculated with Minuit's MINOS method, are shown with semi-transparent color and thinner lines.} 
\label{fig:lat_seds}
\end{figure*}

\section{Modeling of the multi-wavelength emission}\label{sec:method}

\subsection{Theoretical framework}\label{sec:model}

If we assume that the bulk of the flaring activity in blazars originates from large instabilities at the base of the jet, where external radiation fields are more intense, most leptons lose their energy in the production of soft $\gamma$-rays through inverse Compton scattering on external photon fields (also known as external Compton, EC). Consequently, the production of synchrotron photons is interrupted at relatively low energies \citep[e.g.,][]{cost}. 
In the absence of these additional photon fields, e.g. with the so called ``naked'' AGN structure, a black hole is only surrounded by weak accretion flows and low power jets. Compton cooling losses are therefore less severe and electrons can potentially gain more energy. 
The resulting synchrotron emission peaks in the UV to X-ray band and provides the only internal radiation field that can interact through Compton scattering with the same population of ultra-relativistic leptons \citep[e.g.,][]{cost}. The emission of this synchrotron-self-Compton (SSC) radiation sometimes reaches up to TeV energies. The Cosmic Microwave Background (CMB), while theoretically a source of photons for EC scattering for any jet, is negligible because its intensity scales with $(1+z)^4$ and all our blazars have relatively low redshifts \citep[][and references therein]{2020Galax...8...72C}. 

This simple scenario would naturally explain why the synchrotron peak in FSRQs, LSPs and ISPs sits at lower frequencies than in (E)HSP blazars and why the latter represent the most numerous class of extragalactic TeV energy emitters \citep[][see also TeVCat\footnote{TeVCat \protect\url{http://tevcat.uchicago.edu/} is an online interactive catalog of sources historically detected at $E > 0.1$ TeV. The catalog reports approximately 90 extragalactic sources as of November 2021. Among them, 55 are HSPs.}]{horan}.
It must be remarked however that this classification is not fixed and transitions or ``identity crises'' have been spotted for a number of AGN \citep[][]{tavecchio2001,emmanoulopoulos2012,ahnen2015}. 

In the classical picture, the broadband spectrum of HSPs is dominated by non-thermal emission. Thermal components, if present, are often associated with old star populations from the giant elliptical host galaxy.
The broad and narrow line regions (BLR and NLR), are either absent or very dim for these objects, and are expected to be subdominant compared to the SSC radiation. The same applies to the disk and dusty torus emission. An example of a HSP blazar is Markarian 501 \citep[e.g.][]{abdo2011}. 

\subsubsection{Source geometry}

The geometry of the emitter, including the orientation and position of its individual components with respect to the observer, has important implications in the production of ultra-relativistic particles and the emission of $\gamma$-rays. Because of the limited data available, we assume one of the simplest source geometries that we can think of. The blazar structure consists of a central black hole (SMBH) with a plasma of ultra-relativistic electrons at a distance from the SMBH of $\mathrm{R_H}$, likely accelerated in a shocked medium. This plasma is embedded inside a luminous jet closely aligned with the line-of-sight, and accompanied with a tangled magnetic field of strength $\mathrm{B}$.
On first approximation, we assumed that the accretion disk and broad or narrow line regions of all our sources are either absent or too weak to be detected, assumption supported by the lack of a relatively weak emission of the AGN structure in the optical band compared to the thermal host component. We however allow for a dusty torus component to provide an additional external Compton process in a few sources, which could explain both the pronounced $\gamma$-ray spectral curvature and hints of an additional IR component apparent in some of the archival photometric data.

\subsubsection{Synchrotron and inverse Compton emission}

In this scheme, the bulk of the radio through hard X-ray emission from the AGN structure would be explained as synchrotron radiation from a population of ultra-relativistic electrons embedded in the jet. The spectrum of these particles is assumed to have a broken power-law shape with low-energy spectral index $p_1$ between Lorentz factors $\gamma_\mathrm{min}$ and $\gamma_\mathrm{br}$, and high-energy spectral index $p_2$ between $\gamma_\mathrm{br}$ and $\gamma_\mathrm{max}$. At radio frequencies, the scattering between synchrotron photons and electrons induces the so called synchrotron self-absorption (SSA), a suppression factor that occurs at low frequencies, at which the mean free path is small compared to the size of the emitting region. The high energy component, covering energies from  below approximately $1\,\mathrm{MeV}$ to several TeV, is explained as inverse Compton scattering of the same population of ultra-relativistic electrons on both the synchrotron radiation and, if present, the external infrared radiation field from the torus.

\subsubsection{Host emission}

Following \cite{donea2003}, we included an additional black body emission component to simulate, on first approximation, the contribution of the host galaxy to the observed SED. Even though we only consider the host as a single-temperature thermal emitter with effective temperature $\mathrm{T_{eff,host}}$ and luminosity $\mathrm{L_{host}}$, we find the resulting modeling accurate enough for our purposes, assuming that the host is indeed a passive evolving  giant elliptical galaxy. 
If needed, the actual shape could in principle be reproduced in future works through stellar synthesis models.
Because the host emission is external to the blazar structure, we do not expect it to increase significantly the amount of seed photons at the emission region. Therefore, we neglect its external Compton contribution at high energies.

\subsection{Broadband spectral energy distribution modeling}

\subsubsection{Fitting}\label{sec:modelfit}

We modelled the broadband SEDs for the 22 selected sources using the JetSet package \citep{massaro2006,tramacere2009,tramacere2011}, including all the data described in sections \ref{sec:mw-archival} and \ref{sec:obs_analysis}. We used JetSet's fitting capabilities to estimate the broadband spectral model parameters, which are based on {\tt Minuit}'s code from \cite{minuit}. 
The high energy {\em Fermi}-LAT spectra used in the modeling was chosen between the LP and the EPWL for each source. Because the models are not nested, we cannot directly compare the likelihood values of each fit to quantify how much one model is preferred over the other. Nevertheless, we can use the Akaike information criterion \citep[AIC,][]{akaike}, an estimator of the prediction error and relative quality of statistical models, to select which model is preferred by the data. In addition to the maximum value of the likelihood function for the model $\hat L$, AIC includes a correction factor to account for the number of estimated model parameters $k$: $AIC = 2k - 2\ln \hat L$. For the particular case of models with the same number of free parameters, the comparison is reduced to the direct comparison of the AIC states that the model providing the highest log-likelihood is the one favored by the data. 
We note that a rigorous analysis would need to take into account spectral data correlation, very relevant for X-ray and $\gamma$-ray instruments, differences in exposure and sensitivity between the different instruments (conditioning the best-fit model to possibly ignore the spectral information from some bands), possible flaring episodes and the homogeneity of the rules of the flux uncertainty estimation for all involved wavelength bands. These issues are often impossible to correctly address using archival data that in many cases consist of flux data with their error bars, which are estimated using different methodology.

\subsubsection{Model assumptions and observational constraints}

The proposed model provides a good representation of the observed spectral energy distributions.
We are well aware that the available data are not sufficient to strongly constrain any complex model for most sources. However, it is enough to study which model parameters are most sensitive to the different wavelength bands and to discuss observation strategies and new data required to improve the constraining power on the proposed models.

To break the large parameter degeneracy in the model, we fixed some parameters to typical values that we found in the literature \citep[see e.g.][where the SSC/EC modeling of blazars and the resulting parameters are discussed in detail]{2011ApJ...738..157Z,2016MNRAS.456.2374T, 2018A&A...616A..63A}. 
The radius of the emitting region was set to $R=10^{16}\,\mathrm{cm}$ and its distance placed at $R_H = 2\times 10^{18}\,\mathrm{cm}$. Given the lack of constraining data in the far-IR, we tested two possible, very different, values of the minimum Lorentz factor: $\gamma_{\rm min}=1$ and $\gamma_{\rm min} = 1\times 10^3$. The bulk Lorentz factor is also fixed to $\Gamma_{\rm bulk} = 20$. We left free the density of particles $\mathrm{N}$ (normalized so that $\int_{\gamma_{\rm min}}^{\gamma_{\rm max}} n(\gamma)d\gamma \equiv 1$) and magnetic field strength $\mathrm{B}$. Both parameters govern the balance between the fluxes emitted in the low and high energy components and have an impact on the blazar bolometric luminosity.

\begin{figure*}
\centering
\includegraphics[width=0.325\linewidth]{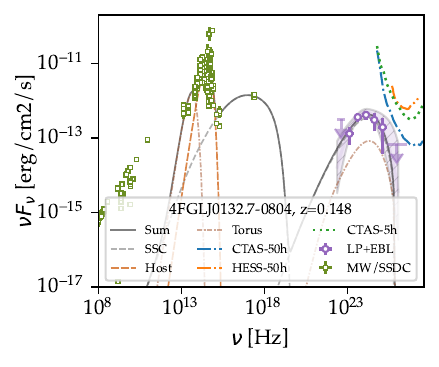}
\includegraphics[width=0.325\linewidth]{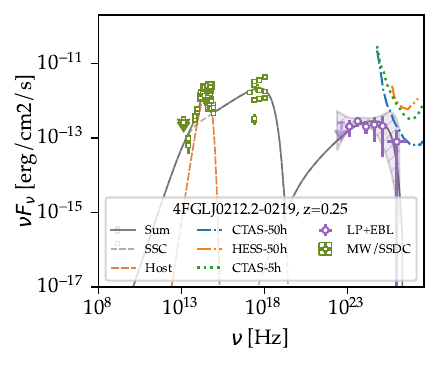}
\includegraphics[width=0.325\linewidth]{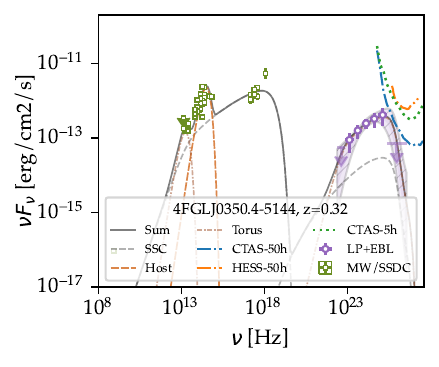}
\includegraphics[width=0.325\linewidth]{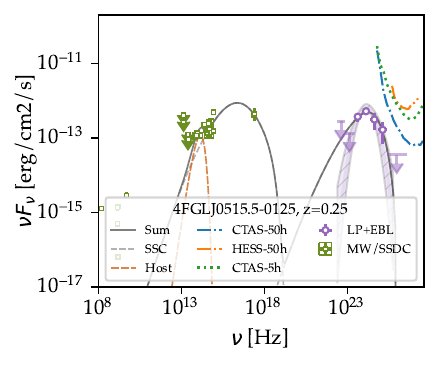}
\includegraphics[width=0.325\linewidth]{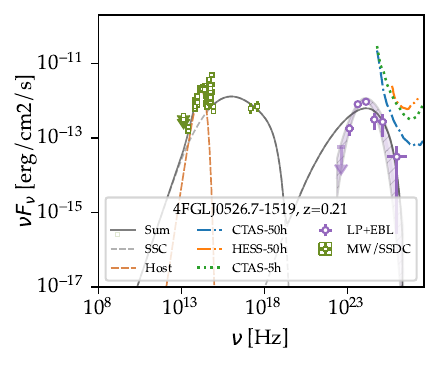}
\includegraphics[width=0.325\linewidth]{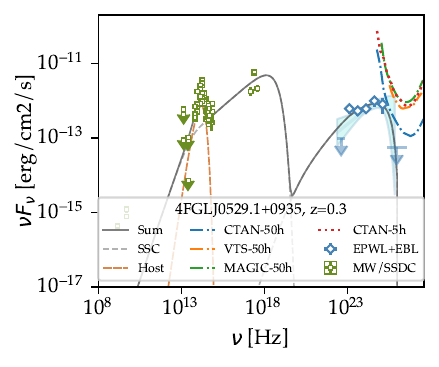}
\includegraphics[width=0.325\linewidth]{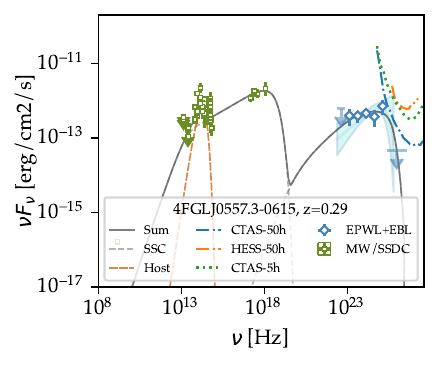}
\includegraphics[width=0.325\linewidth]{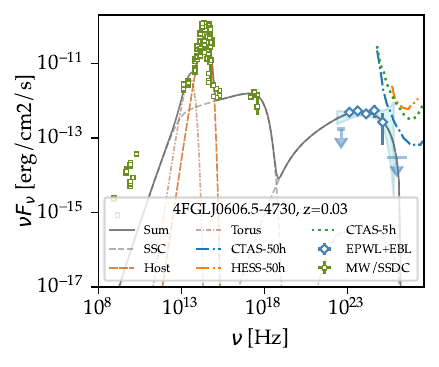}
\includegraphics[width=0.325\linewidth]{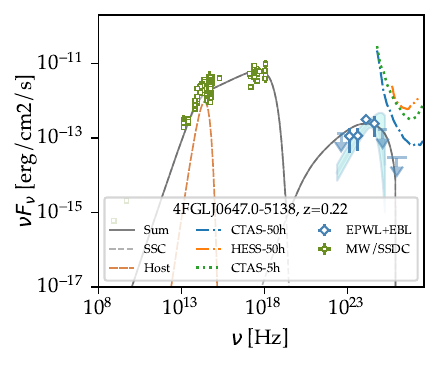}
\includegraphics[width=0.325\linewidth]{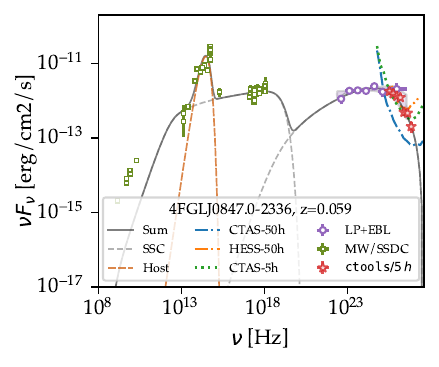}
\includegraphics[width=0.325\linewidth]{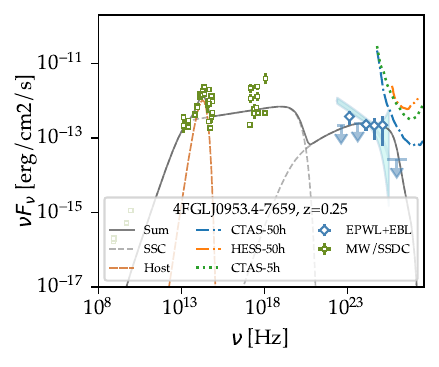}
\includegraphics[width=0.325\linewidth]{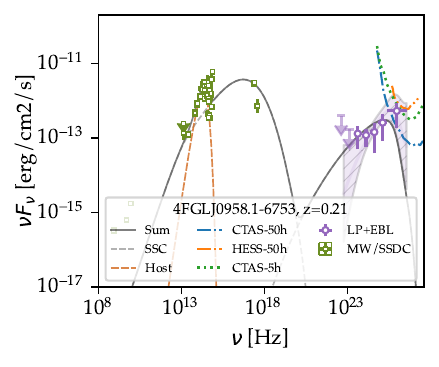}
\includegraphics[width=0.325\linewidth]{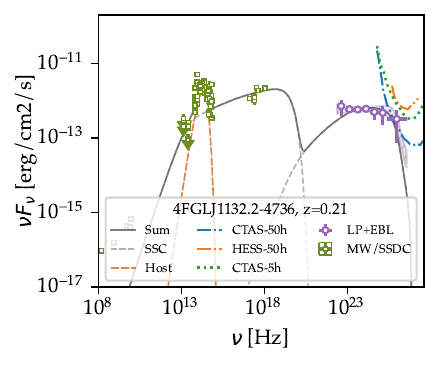}
\includegraphics[width=0.325\linewidth]{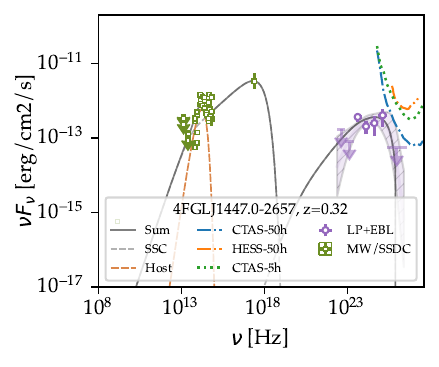}
\includegraphics[width=0.325\linewidth]{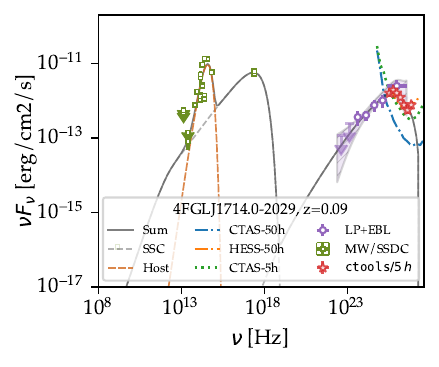}
\label{fig:mw_seds0}
\end{figure*}

\begin{figure*}
\ContinuedFloat
\centering
\includegraphics[width=0.325\linewidth]{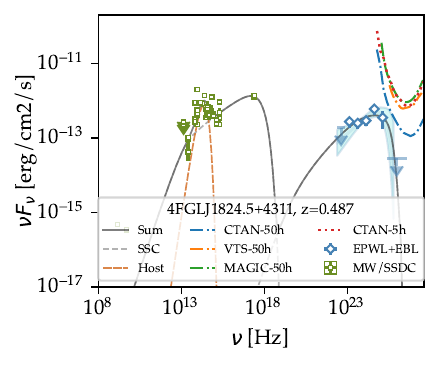}
\includegraphics[width=0.325\linewidth]{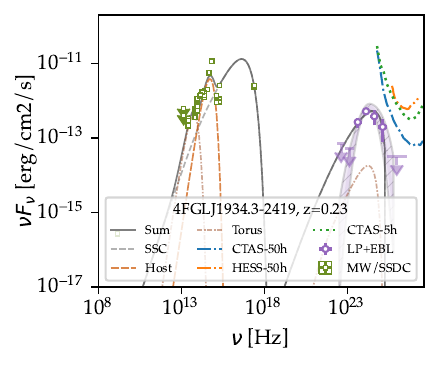}
\includegraphics[width=0.325\linewidth]{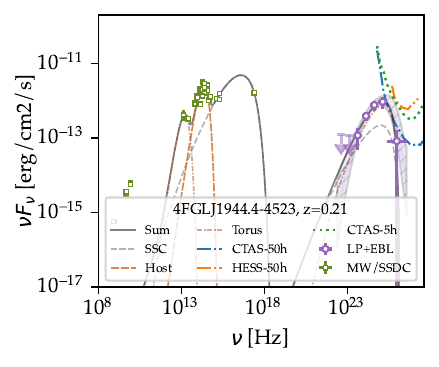}
\includegraphics[width=0.325\linewidth]{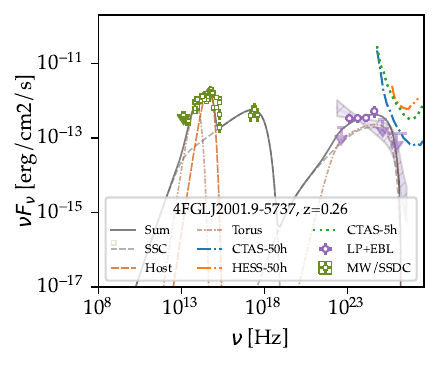}
\includegraphics[width=0.325\linewidth]{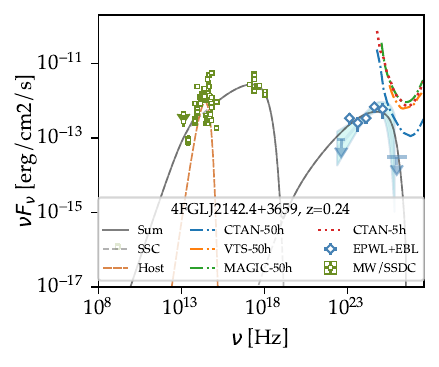}
\includegraphics[width=0.325\linewidth]{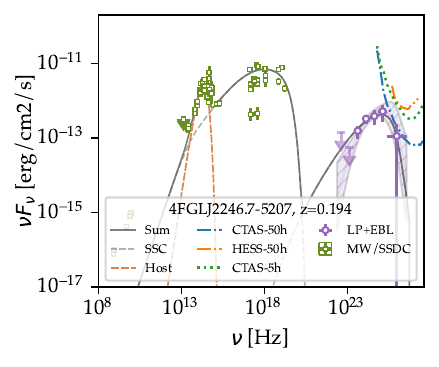}
\includegraphics[width=0.325\linewidth]{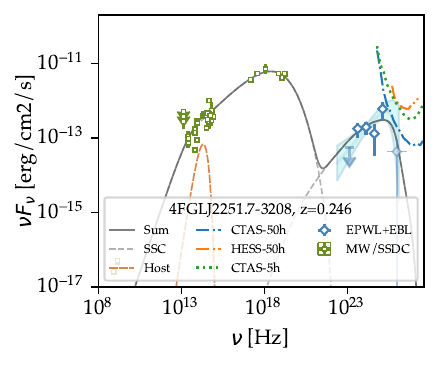}
\caption{Multi-wavelength emission of the selected sources, including archival measurements, the re-analysis of HE $\gamma$-ray data from {\em Fermi}-LAT and `toy' Monte Carlo based SED reconstructed with {\tt ctools} from the Prod-3b CTA simulations, described in section \ref{sec:vhe_prospects}. In all cases, the one-zone leptonic emission model, with fixed minimum Lorentz factor $\gamma_{\rm min} = 1\times10^3$, plus thermal components (host emission and possibly a dusty torus) has been fitted to the measured points (excluding {\tt ctools} points). 
The model include the effects from EBL absorption to reproduce the observed (EBL-absorbed) $\gamma$-ray data.}
\label{fig:mw_seds}
\end{figure*}

We note that, despite our efforts to reduce the number of free parameters, the proposed model has in its simplest form nine free parameters, increasing to 12 for the sources with an assumed dusty torus. As we mentioned, some are degenerate given the limited coverage of the broadband spectra that we have. For example, larger photon fluxes in the high energy component could be obtained either by increasing the particle density $N$, having a slightly harder electron spectral index $p_1/p_2$ or varying the jet angle $\theta$. Distinguishing between the three effects would require in many cases to have significantly better spectral coverage accompanied with more precise flux measurements in most of the bands.

\subsection{Results}

\subsubsection{Broadband emission}
Figure \ref{fig:mw_seds} shows the broadband SEDs collected for the 22 selected targets, including the archival data and the re-analysis of $\gamma$-ray data from {\em Fermi}-LAT using the spectral shape selected with the AIC, as presented in section \ref{fig:lat_seds}. Using the prescriptions from section \ref{sec:model}, we successfully reproduced the different SEDs. The total emission model, together with the contributions from the different components (SSC, host emission and torus), is also shown in this figure. 

The best-fit parameters are summarized in Table \ref{tab:jet_parameters}. 
The characteristics of the electron spectrum are reasonably well constrained by the observed data, with the exception of the total density of electrons $N$. In general, {\em Fermi}-LAT gives us a good handle on the spectral index of the electrons at low energies, while the sub-TeV part of the {\em Fermi}-LAT data, together with the measurements in the UV to X-rays gives us a somewhat accurate depiction of the spectral index of electrons at high energies. 

The lack of photometric measurements at sub-mm and far-IR/mid-IR results in very loose constraints on  the position of the break $\gamma_{\rm break}$, which can only be set from {\em Fermi}-LAT measurements. 
The Lorentz factor range of the electrons, given by $\gamma_{\rm min}$ and $\gamma_{\rm max}$ could not be effectively determined with precision in any case because of the small amount of data available at hard X-rays and soft $\gamma$-rays. We conservatively constrained $\gamma_{\rm max}$ using mainly X-ray together with sub-TeV $\gamma$-ray observations. For the two values of the minimum Lorentz factor tested, we observe that higher values of that parameter, $\gamma_{\rm min}=1\times 10^3$, result in a better reproduction of the $\gamma$-ray data for several sources, particularly at energies below $\sim 1\,\mathrm{GeV}$. As a result, we focus on that solution in this section, leaving the discussion of the modeling with $\gamma_{\rm min}=1$ for the appendix \ref{sec:gammamin1}.

The Doppler boosting factor $\delta$ is set to vary freely during the fit by leaving free the viewing angle $\theta$. We find reasonably low $\delta$ values, with only a handful cases with values of up to $40$ (the maximum allowed in the fit, corresponding to jets fully aligned with the line-of-sight). For the viewing angle, we obtained values of $\lesssim 10^\circ$ for most sources, in line with their classification as blazars.

Under the assumption of a simple black-body host emission component, we find host temperatures of the order of $(2-5)\times10^3\,\mathrm{K}$ for most sources, which would correspond to a population of old stars, in agreement with our expectations \citep[e.g.][]{2000ApJ...532..816U}. Notable exceptions include J1934.3-2419 and J2001.9-5737, which have temperatures of this black body emitter of about $6-8\times 10^3\,\mathrm{K}$. This could point towards non-negligible emission from an accretion disk, perhaps partly obscured.  
For specific sources, we observe a hint of excess in the archival IR data, which cannot be described with our simplified host emission model; or very curved $\gamma$-ray spectra, difficult to reproduce unless unrealistically low values of $p_1$ are used. The latter would have implications on the shape of the synchrotron emitted spectrum. In such cases, we failed to get a reasonable model fit using a pure SSC model to the complete dataset. To improve the reproduction of the observed data, we added a putative IR torus with its associated inverse Compton emission. Its temperature and radius are left free, obtaining in most cases temperatures of a few tens to hundreds of K. Only in two objects, J1934.3-2419 and J2001.9-5737, we found an IR excess with a temperature around $10^3\,\mathrm{K}$, suggesting either the presence of a more complicated thermal component and the existence of an effective nuclear heating source or that the emission is too complex to be successfully reproduced using our simplistic black body host template.

\subsubsection{EHSP candidates}

\begin{figure}
\centering
\hspace*{-5pt}\includegraphics[width=0.48\textwidth]{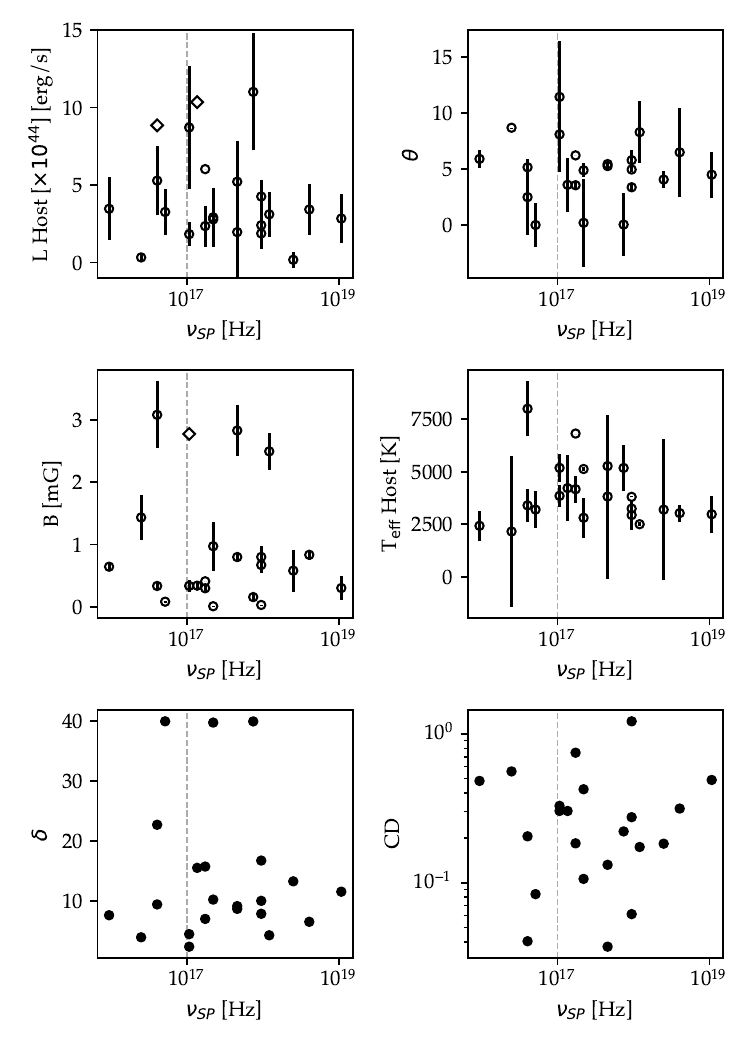}
\caption{
Best fit parameters (from first row left to second row right: host luminosity, jet angle, magnetic field strength and effective temperature of the host galaxy) as a function of the calculated synchrotron peak frequency. Filled circles show parameter values constrained by the modeling and open diamonds represent parameters not constrained by the model
(statistical uncertainty greater than $50\%$ of the maximum parameter value for the full source sample). Last row: Doppler factor (left) and Compton dominance (right) as a function of the calculated synchrotron peak frequency. The vertical gray dashed line separates HSPs from EHSPs.
}
\label{fig:par_correlations}
\end{figure}

One of the main advantages of the modeling technique presented in section \ref{sec:modelfit} is the direct estimation of the synchrotron peak location, hidden otherwise by thermal components for some sources. The last two columns of Table \ref{tab:model_parameters} represent the frequency (in log-scale) and the spectral classification of the blazars presented in this work. Out of the 22 sources, 17 are classified as EHSPs and five as HSPs. From the EHSPs, four sources stand out with extreme synchrotron peak frequencies, i.e. $\mathrm{\nu_{SP}} > 10^{18}\,\mathrm{Hz}$. They are J0529.1+0935, 
J0953.4-7659, J1132.2-4736, and J2251.7-3208. Only 
J2251.7-3208 has enough X-ray data to directly constrain the peak location, while for the others the quoted value represents a best-guess derived mostly from {\em Fermi}-LAT data. 
As seen in Figure \ref{fig:par_correlations}, we did not find a clear correlation between the synchrotron peak frequencies and the model parameters, except for the obvious connection to the underlying spectral parameters of the electrons.

\subsection{Gamma-ray variability implications}

An important caveat of this analysis is the assumption that our sources are not violently variable emitters. This allowed us to include archival data not simultaneous with the $\gamma$-ray observations. We note however that variability can be energy dependent, and more importantly, {\em Fermi}-LAT's limited instantaneous sensitivity at GeV energies poses important limitations in the determination of variability for HSPs and EHSPs. Many HSP and EHSP do not show large scale variability in $\gamma$-rays, but are clearly variable in the X-ray band. 
To test the no-variability assumption in $\gamma$-rays, we extracted the variability index and the fractional variability from the 4FGL-DR2 catalog (Table \ref{tab:variability}) and compared them to other sources in that catalog. Variability index is defined as the sum of the log-likelihood difference between the flux fitted in each 4FGL time interval and the average flux over the full 4FGL dataset. A value greater $18.5$ over 12 intervals implies a high chance ($>99\%$) of being a variable source. Only one source in our sample (4FGL~J1714.0-2029) is clearly variable according to this test. 
We also considered the fractional variability $F_\mathrm{var}$. $F_\mathrm{var}$ is defined in \citet{4fgl} as

\begin{equation}
    \frac{\delta F}{F_{\rm av}} = \frac{\sqrt{\max\left[\frac{1}{N-1}\sum_i (F_i - F_{\rm av})^2-\frac{\sum_i \sigma_i^2}{N_{\rm int}},0\right]}}{F_{\rm av}}
\end{equation}

\noindent where $F_{\rm av}$ is the average flux, $F_i$ the individual flux measurements, $N_{\rm int}$ the number of bins and $\sigma_i$ the statistical error.
$F_\mathrm{\rm var}$ in HE is small and even compatible with $0$ for most cases, in contrast with highly variable sources, for which values of $F_\mathrm{\rm var}>1$ are not uncommon. J1714.0-2029, with $F_\mathrm{\rm var}=0.69\pm0.31$ (value higher than that of about $80\%$ of the sources of the 4LAC-DR2), is classified as the most variable source in the sample.

\begin{figure}
\centering
\includegraphics[height=0.75\linewidth]{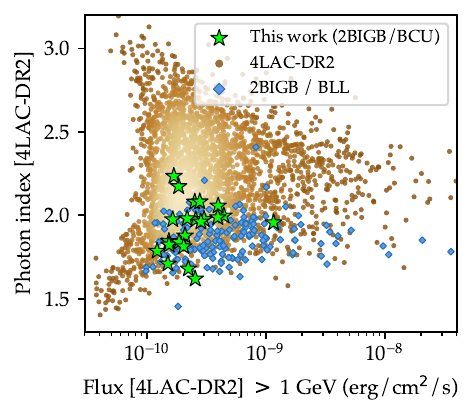}
\includegraphics[height=0.75\linewidth]{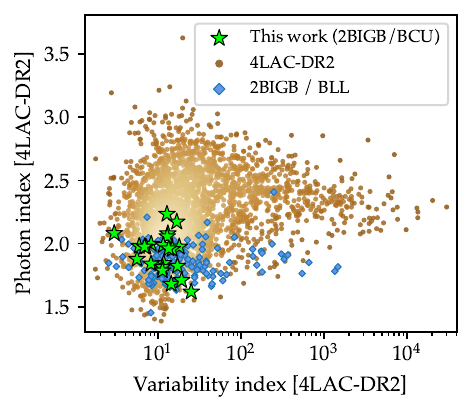}
\includegraphics[height=0.75\linewidth]{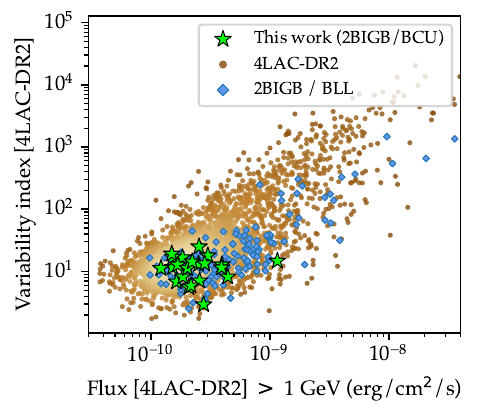}
\caption{{\bf Top:} Photon spectral index as a function of flux; {\bf Middle:} Photon spectral index as a function of variability index in the {\em Fermi}-LAT band; {\bf Bottom:} Variability index as a function of flux. In brown circles we show all the sources from the 4LAC-DR2 catalog, including those of unknown type. Density is color coded, with lighter brown representing larger density of sources. We represent in blue confirmed BLL from the 2BIGB with $\mathrm{TS_{2BIGB}}>25$ and a significance of at least $5\,\sigma$ in the 4FGL, good MWL coverage ({\tt sflag=0}), measured synchrotron peak $\mathrm{\nu_{SP}}$ or at least a lower limit ({\tt nuflag $\in$ [1,3])}, FOM of at least 0.7 and a firm spectroscopic or estimated photometric redshift, either with or without features in the optical band ({\tt zflag $\in$ [1,4,5]}). In green stars we plot blazars of unknown type in the 4LAC-DR2 which have passed the aforementioned cuts. }
\label{fig:sample_context}
\end{figure}

Figure \ref{fig:sample_context} shows the dependency of the spectral index of the 2BIGB sources on the flux and the variability index and the dependency between the latter two parameters for different selection criteria (BLL, BCUs and the 22 selected sources). As can be seen, 2BIGB BLL and BCUs tend to display hard spectra and weak variability when compared to the general population of blazars in the LAC-DR2 (which contains also FSRQs and LSPs/ISPs). In particular, we note that the selected BCU are found preferably at low fluxes compared to 2BIGB BLL. They also have, for a given flux level, lower variability indices compared to the general population of sources in the 4LAC-DR2.

\begin{figure}
\centering
\includegraphics[width=0.48\textwidth]{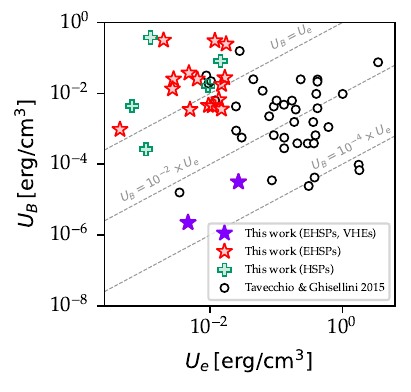}
\caption{
Magnetic energy density as a function of kinetic energy density derived from our sample (red and purple stars for sources classified as EHSP and green crosses for HSP classified objects) and the one for BLLs presented in \protect\cite{2016MNRAS.456.2374T} (open black circles). 
Dashed  gray lines show the $U_B = U_e$ (equipartition), $U_B=10^{-2} \times U_e$ and $U_B=10^{-4} \times U_e$ cases. 
Our sample has, on average, slightly more magnetized jets for a given kinetic energy that the sample of BLLs presented in \protect\cite{2016MNRAS.456.2374T}, and closer in general to equipartition (except for the two VHE candidates).
}
\label{fig:equipartition}
\end{figure}

Finally, we explored the balance between rest-frame magnetic and kinetic energy density for our sample of blazars in Figure \ref{fig:equipartition}, distinguishing between those classified as HSPs and EHSPs. Compared to the selection in \cite{2016MNRAS.456.2374T}, which predominantly cluster around $U_B/U_e \sim 10^{-2}$, our sources tend to have slightly higher values of such ratio, even though we find two sources still far from equipartition: J0847.0-2336 and J1714.0-2029.

\begin{landscape}
\begin{table}
\caption{Main jet model best-fit parameters, including electron density $\mathrm{N}$, 
magnetic field strength $\mathrm{B}$, 
electron indices before and after the break $p_1$ and $p_2$, 
maximum Lorentz factors $\gamma_\mathrm{max}$ and position of the spectral break $\gamma_\mathrm{br}$, 
jet angle with respect to the line of sight $\theta$, 
Doppler boosting $\delta$, 
estimated position of the synchrotron peak $\nu_\mathrm{SP}$, 
favored LAT model to reproduce the high energy data: Log Parabola (LP) or Power Law with exponential cut-off (EPWL),  
Compton dominance (CD) and the classification as either HSP or EHSP. 
Fixed values: radius of the emitting region $\mathrm{R}=1.0\times 10^{16}\, \mathrm{cm}$, position with respect to the central object $\mathrm{R_H}=2.0\times 10^{18}\, \mathrm{cm}$. Bulk Lorentz factor is fixed to $\Gamma=20$.} Assumed minimum Lorentz factor $\gamma_{\rm min}=10^3$. [$\dagger$] denotes unconstrained best-fit parameters.
\centering
\label{tab:jet_parameters}
\begin{tabular}{lrrrrrrrrrrrr}
\toprule
4FGL Name & $\mathrm{N}$ & $\mathrm{B}$ & $p_1$ & $p_2$ & $\gamma_\mathrm{br}$ & $\gamma_\mathrm{max}$ & $\theta$ & $\delta$ & $\log{\mathrm{\nu_{SP}}}$ & LAT model & CD & Class\\
& & & & & $[\times 10^{4}]$ & $[\times 10^{6}]$ & & & & & \\
& $\mathrm{(cm^{-3})}$ & $\mathrm{(G)}$ & & & & & $\mathrm{(deg})$ & & $\mathrm{[Hz]}$ & & \\
\midrule
J0132.7$-$0804 &               $0.04 [\dagger]$ &           $2.8\pm1.6$ &       $0.65\pm0.08$ &         $4.0\pm2.5$ &                  $3.0\pm1.6$ &             $0.35\pm0.21$ &              $11\pm5$ & $2.4$ &     $17.0$ &        LP &   $0.3$ &  EHSP \\
J0212.2$-$0219 &     $(6.2\pm0.9)\times 10^{4}$ &         $0.80\pm0.08$ &     $2.413\pm0.013$ &         $4.2\pm1.6$ & $0.4\times 10^{3} [\dagger]$ &             $0.30\pm0.05$ &           $5.4\pm0.4$ & $8.7$ &     $17.7$ &        LP &  $0.13$ &  EHSP \\
J0350.4$-$5144 &         $(5\pm4)\times 10^{1}$ &         $0.15\pm0.06$ &     $1.831\pm0.015$ &       $2.55\pm0.21$ &              $0.5 [\dagger]$ &             $0.40\pm0.20$ &       $0.1 [\dagger]$ &  $40$ &     $17.9$ &        LP &  $0.22$ &  EHSP \\
J0515.5$-$0125 &                $0.006\pm0.004$ &           $1.4\pm0.4$ &       $0.38\pm0.07$ &         $6.0\pm0.9$ &                  $4.4\pm1.4$ &         $0.2841\pm0.0024$ &       $8.7 [\dagger]$ & $3.9$ &     $16.4$ &        LP &  $0.56$ &   HSP \\
J0526.7$-$1519 &              $0.001 [\dagger]$ &         $0.64\pm0.07$ &    $0.06 [\dagger]$ &         $4.0\pm0.4$ &                $0.87\pm0.32$ &             $0.40\pm0.22$ &           $5.9\pm0.8$ & $7.6$ &     $16.0$ &        LP &  $0.48$ &   HSP \\
J0529.1$+$0935 &   $0.6\times 10^{4} [\dagger]$ &         $2.49\pm0.30$ &         $1.9\pm0.5$ &         $2.2\pm0.7$ &             $0.01 [\dagger]$ &             $0.36\pm0.10$ &           $8.3\pm2.8$ & $4.3$ &     $18.1$ &      EPWL &  $0.17$ &  EHSP \\
J0557.3$-$0615 &     $(2.5\pm0.7)\times 10^{5}$ &         $0.67\pm0.12$ &     $2.477\pm0.026$ &         $2.5\pm1.3$ &   $(2.1\pm1.7)\times 10^{3}$ &           $0.5 [\dagger]$ &           $5.8\pm0.9$ & $7.9$ &     $18.0$ &      EPWL &  $0.28$ &  EHSP \\
J0606.5$-$4730 &     $1\times 10^{4} [\dagger]$ &         $0.33\pm0.10$ &         $1.9\pm0.5$ &         $2.8\pm0.8$ &             $0.05 [\dagger]$ &             $0.38\pm0.05$ &           $8.1\pm3.4$ & $4.5$ &     $17.0$ &      EPWL &  $0.33$ &  EHSP \\
J0647.0$-$5138 &     $(3.1\pm0.9)\times 10^{3}$ &           $2.8\pm0.4$ &     $1.948\pm0.031$ &       $2.59\pm0.07$ &              $0.0 [\dagger]$ & $0.155 [\dagger]$ &   $5.27 [\dagger]$ & $9.1$ &     $17.7$ &      EPWL & $0.037$ &  EHSP \\
J0847.0$-$2336 &   $1.3\times 10^{4} [\dagger]$ &     $0.0281\pm0.0020$ &       $1.90\pm0.18$ &       $2.86\pm0.26$ &             $0.18 [\dagger]$ &               $2.4\pm1.5$ &           $3.4\pm0.4$ &  $17$ &     $18.0$ &        LP &   $1.2$ &  EHSP \\
J0953.4$-$7659 &     $(5.8\pm2.0)\times 10^{6}$ &         $0.30\pm0.19$ &       $2.86\pm0.07$ &         $2.9\pm2.8$ & $0.9\times 10^{3} [\dagger]$ &             $3.00\pm0.24$ &           $4.5\pm2.0$ &  $12$ &     $19.0$ &      EPWL &  $0.49$ &  EHSP \\
J0958.1$-$6753 &     $1\times 10^{1} [\dagger]$ &       $0.081\pm0.016$ &         $1.6\pm0.5$ &            $10\pm4$ &                    $38\pm32$ &             $2.68\pm0.22$ &       $0.0 [\dagger]$ &  $40$ &     $16.7$ &        LP & $0.084$ &   HSP \\
J1132.2$-$4736 &     $(2.9\pm2.7)\times 10^{5}$ &         $0.83\pm0.08$ &     $2.468\pm0.016$ &       $2.80\pm0.12$ &                $2 [\dagger]$ &             $1.35\pm0.07$ &               $6\pm4$ & $6.5$ &     $18.6$ &        LP &  $0.31$ &  EHSP \\
J1447.0$-$2657 &   $0.6\times 10^{4} [\dagger]$ &           $1.0\pm0.4$ &       $2.17\pm0.17$ &         $4.3\pm1.9$ &   $(3.2\pm2.2)\times 10^{3}$ &     $0.145 [\dagger]$ &           $4.9\pm0.6$ &  $10$ &     $17.3$ &        LP &  $0.11$ &  EHSP \\
J1714.0$-$2029 &                     $123\pm27$ &     $0.007 [\dagger]$ &     $1.810\pm0.020$ &         $4.0\pm1.4$ & $0.5\times 10^{3} [\dagger]$ &           $0.7 [\dagger]$ &         $0.0 [\dagger]$ &  $40$ &     $17.3$ &        LP &  $0.42$ &  EHSP \\
J1824.5$+$4311 &     $(2.5\pm1.9)\times 10^{3}$ &         $0.34\pm0.07$ &       $1.92\pm0.09$ &       $2.41\pm0.21$ &              $0.3 [\dagger]$ &           $0.2 [\dagger]$ &           $3.6\pm2.4$ &  $16$ &     $17.1$ &      EPWL &   $0.3$ &  EHSP \\
J1934.3$-$2419 &               $0.05 [\dagger]$ &           $3.1\pm0.5$ &       $0.96\pm0.19$ &       $2 [\dagger]$ &                $7 [\dagger]$ &         $0.0295\pm0.0027$ &           $5.2\pm0.4$ & $9.4$ &     $16.6$ &        LP &  $0.04$ &   HSP \\
J1944.4$-$4523 &             $1.2\times10^{-4} [\dagger]$ &         $0.33\pm0.05$ &     $0.337\pm0.033$ &             $6\pm4$ &                  $5.6\pm0.8$ &          $0.08 [\dagger]$ &       $2.5 [\dagger]$ &  $23$ &     $16.6$ &        LP &   $0.2$ &   HSP \\
J2001.9$-$5737 & $6.323 \times 10^{4} [\dagger]$ & $0.408 [\dagger]$ & $2.33 [\dagger]$ & $2.48 [\dagger]$ &           $20.85 [\dagger]$ &   $0.276 [\dagger]$ & $6.23 [\dagger]$ & $7.0$ &     $17.2$ &        LP &  $0.75$ &  EHSP \\
J2142.4$+$3659 &     $(4.2\pm1.1)\times 10^{3}$ &         $0.30\pm0.05$ &     $2.115\pm0.021$ &         $3.1\pm0.6$ &                      $7\pm4$ &             $0.27\pm0.24$ &         $3.56\pm0.29$ &  $16$ &     $17.2$ &      EPWL &  $0.18$ &  EHSP \\
J2246.7$-$5207 &                       $54\pm6$ &         $0.80\pm0.18$ &     $1.702\pm0.011$ &       $3.73\pm0.15$ &                 $12.7\pm1.6$ &             $0.85\pm0.11$ &         $4.96\pm0.21$ &  $10$ &     $18.0$ &        LP & $0.061$ &  EHSP \\
J2251.7$-$3208 &     $(2.3\pm1.0)\times 10^{3}$ &         $0.58\pm0.34$ &       $2.14\pm0.05$ &        $11.9\pm2.3$ &   $(3.9\pm2.2)\times 10^{2}$ &               $7.5\pm1.1$ &           $4.1\pm0.7$ &  $13$ &     $18.4$ &      EPWL &  $0.18$ &  EHSP \\
\bottomrule
\end{tabular}
\end{table}
\end{landscape}

\begin{landscape}
\begin{table}
\centering
\caption{Energy budget, showing the effective temperature $\mathrm{T_{eff,host}}$ of the black body that we added to simulate host galaxy emission, 
the effective temperature of the dusty torus $\mathrm{T_{DT}}$, 
the integrated host luminosity $\mathrm{L_{host}}$ and the luminosity carried by the jet for the non-thermal low energy and total radiative components $\mathrm{L_{sync}}$ and $\mathrm{L_{rad}}$, 
the electrons $\mathrm{L_{kin}}$, the Poynting luminosity due to the magnetic field} $\mathrm{L_B}$, 
the total jet luminosity $\mathrm{L_{tot}}$ and the energy density ratio of the magnetic field to that of the electron distribution $\mathrm{U_B/U_e}$.
$[\dagger]$ denotes unconstrained best-fit parameters.
\label{tab:model_parameters}
\begin{tabular}{lrrrrrrrrrrr}
\toprule
4FGL Name  & $\mathrm{R_{DT}}$ & $\mathrm{T_{DT}}$ & $\tau_\mathrm{DT}$ & $\mathrm{T_{host}}$ & $\mathrm{L_{host}}$ & $\mathrm{L_{sync}}$ & $\mathrm{L_{rad}}$ & $\mathrm{L_{B}}$ & $\mathrm{L_{kin}}$ & $\mathrm{L_{tot}}$ & $\mathrm{U_B/U_e}$ \\
& $[\times 10^{18}]$ & $[\times 10^{2}]$ & & $[\times 10^{3}]$ & $[\times 10^{44}]$ & $[\times 10^{42}]$ & $[\times 10^{42}]$ & $[\times 10^{42}]$ & $[\times 10^{44}]$ & $[\times 10^{44}]$ & \\
& $\mathrm{(cm)}$ & $\mathrm{(K)}$ & &  $\mathrm{(K)}$ & $\mathrm{(erg/s)}$ & $\mathrm{(erg/s)}$ & $\mathrm{(erg/s)}$ & $\mathrm{(erg/s)}$ & $\mathrm{(erg/s)}$ & $\mathrm{(erg/s)}$ & \\
\midrule
J0132.7$-$0804 &            $2.0 [\dagger]$ &          $6.1\pm2.3$ &        $0.15 [\dagger]$ &           $5.2\pm0.7$ &             $9\pm4$ & $1.7\times10^{3}$ &  $1.1\times10^{3}$ & $2.0\times10^{3}$ &    $0.44$ &      $33$ &               $26$ \\
J0212.2$-$0219 &                              - &                    - &                       - &           $5.3\pm1.4$ &         $5.2\pm2.6$ &              $46$ &               $95$ &              $52$ &    $0.24$ &     $1.7$ &              $4.0$ \\
J0350.4$-$5144 &                   $13.1\pm1.9$ &          $2.9\pm1.3$ &         $0.063\pm0.017$ &           $5.2\pm1.1$ &            $11\pm4$ &            $0.16$ &              $3.6$ &            $0.19$ &   $0.016$ &   $0.054$ &              $2.2$ \\
J0515.5$-$0125 &                              - &                    - &                       - &         $2 [\dagger]$ &       $0.33\pm0.29$ & $3.8\times10^{2}$ &  $3.0\times10^{2}$ & $5.6\times10^{2}$ &    $0.54$ &     $9.3$ &              $5.7$ \\
J0526.7$-$1519 &                              - &                    - &                       - &           $2.4\pm0.7$ &         $3.5\pm2.0$ &              $35$ &               $62$ &              $49$ &    $0.35$ &     $1.5$ &              $1.8$ \\
J0529.1$+$0935 &                              - &                    - &                       - &         $2.50\pm0.10$ &         $3.1\pm1.4$ & $2.2\times10^{3}$ &  $9.3\times10^{2}$ & $2.6\times10^{3}$ &    $0.65$ &      $37$ &               $14$ \\
J0557.3$-$0615 &                              - &                    - &                       - &           $2.9\pm0.7$ &         $2.4\pm1.5$ &              $86$ &               $67$ & $1.1\times10^{2}$ &    $0.56$ &     $2.4$ &              $1.2$ \\
J0606.5$-$4730 &     $(1.0\pm0.8)\times 10^{2}$ &          $4.3\pm3.1$ &        $0.06 [\dagger]$ &           $3.8\pm0.5$ &         $1.8\pm0.8$ &             $7.4$ &               $17$ &             $9.7$ &     $0.4$ &    $0.67$ &             $0.42$ \\
J0647.0$-$5138 &                              - &                    - &                       - &         $4 [\dagger]$ &     $2.0 [\dagger]$ &              $94$ &  $1.1\times10^{3}$ &              $97$ &   $0.074$ &      $13$ &  $1.6\times10^{2}$ \\
J0847.0$-$2336 &                              - &                    - &                       - &       $3.802\pm0.024$ &         $1.9\pm0.7$ &            $0.22$ &             $0.12$ &            $0.46$ &     $1.0$ &     $1.0$ & $1.1\times10^{-3}$ \\
J0953.4$-$7659 &                              - &                    - &                       - &           $3.0\pm0.9$ &         $2.8\pm1.6$ &             $9.5$ &               $13$ &              $12$ &    $0.56$ &    $0.82$ &             $0.24$ \\
J0958.1$-$6753 &                              - &                    - &                       - &           $3.2\pm0.9$ &         $3.3\pm1.5$ &            $0.12$ &             $0.98$ &            $0.12$ &    $0.04$ &   $0.051$ &             $0.25$ \\
J1132.2$-$4736 &                              - &                    - &                       - &           $3.0\pm0.4$ &         $3.4\pm1.6$ & $1.3\times10^{2}$ &  $1.0\times10^{2}$ & $1.7\times10^{2}$ &    $0.63$ &     $3.4$ &              $1.7$ \\
J1447.0$-$2657 &                              - &                    - &                       - &           $2.8\pm0.9$ &         $2.9\pm1.9$ &              $54$ &  $1.4\times10^{2}$ &              $61$ &    $0.18$ &     $2.2$ &              $7.9$ \\
J1714.0$-$2029 &                              - &                    - &                       - &         $5.13\pm0.09$ &         $2.8\pm0.9$ &           $0.021$ & $8.3\times10^{-3}$ &           $0.031$ &    $0.17$ &    $0.17$ & $4.8\times10^{-4}$ \\
J1824.5$+$4311 &                              - &                    - &                       - &           $4.2\pm1.6$ &            $10\pm6$ &              $12$ &               $17$ &              $16$ &    $0.35$ &    $0.69$ &             $0.49$ \\
J1934.3$-$2419 & $1.4 [\dagger]$ &          $8.4\pm2.3$ &           $0.11\pm0.09$ &           $8.0\pm1.3$ &             $9\pm8$ &              $94$ &  $1.4\times10^{3}$ &              $98$ &   $0.047$ &      $15$ &  $3.0\times10^{2}$ \\
J1944.4$-$4523 &                 $11 [\dagger]$ &        $2 [\dagger]$ &           $0.14\pm0.12$ &           $3.4\pm0.8$ &         $5.3\pm2.2$ &             $1.0$ &               $17$ &             $1.2$ &   $0.025$ &     $0.2$ &              $6.7$ \\
J2001.9$-$5737 &                  $2 [\dagger]$ & $12.3 [\dagger]$ & $0.35 [\dagger]$ & $6.8 [\dagger]$ & $6.0 [\dagger]$ &              $29$ &               $25$ &              $54$ &    $0.51$ &     $1.3$ &             $0.49$ \\
J2142.4$+$3659 &                              - &                    - &                       - &           $4.2\pm0.7$ &         $2.4\pm1.3$ &             $5.1$ &               $13$ &             $6.1$ &    $0.19$ &    $0.38$ &             $0.71$ \\
J2246.7$-$5207 &                              - &                    - &                       - &         $3.24\pm0.04$ &         $4.3\pm1.1$ &              $53$ &               $96$ &              $55$ &     $0.1$ &     $1.6$ &              $9.1$ \\
J2251.7$-$3208 &                              - &                    - &                       - &       $3.2 [\dagger]$ &     $0.2 [\dagger]$ &              $27$ &               $50$ &              $28$ &     $0.1$ &    $0.89$ &              $5.0$ \\
\bottomrule
\end{tabular}
\end{table}
\end{landscape}

\section{Very-high-energy gamma-ray emission}\label{sec:vhe_prospects}

\subsection{Ground-based gamma-ray detectors}

Blazars usually have photon spectral indices of $\alpha>1.5$. As a result, the number of photons from these astrophysical sources that can be detected with space-borne detectors such as {\em Fermi}-LAT (typical collection areas of $\sim 1\,\mathrm{m}^2$) is rather low at energies greater than $100\,\mathrm{GeV}$, unless the source is very luminous or located in the nearby Universe. In addition, $\gamma$-ray photons from extragalactic sources may be subject to EBL absorption, reducing even more the number of photons that can make it to the telescopes as energy increases \citep{magic2019_ebl}.
Instead, VHE emission is usually studied with ground-based instruments, either water Cherenkov detectors \citep[e.g. HAWC, ][]{hawc13} or IACTs such as MAGIC \citep{magic}, VERITAS \citep{veritas}, H.E.S.S. \citep{hess} and the forthcoming Cherenkov Telescope Array \citep[CTA\footnote{\url{https://www.cta-observatory.org/}},][]{2013APh....43....3A}. 
They have effective collection areas which are on the order of $10^5\,\mathrm{m}^2$, much more suited to detect VHE photons. IACTs in particular excel at studying the emission from point-like sources because of their efficient background rejection power and small point spread function (PSF). They are also remarkably good at studying the time-evolving emission from blazars due to their excellent instantaneous sensitivity.

\subsection{Visibility from IACTs}

In order to study the visibility conditions for the best VHE candidate sources with IACTs, we first computed the number of hours for which a given source rises above 40$^{\circ}$ from the local horizon at existing and future IACT sites. 
We set as additional constraints that it is astronomical night time when the observation happens (Sun at least $-18^\circ$ below horizon) and that the Moon is either below the horizon or with a maximum illumination of $50\%$. In addition, because VERITAS does not operate in Summer because of Monsoon conditions present at the Fred Lawrence Whipple Observatory, we excluded the dates between June 15th and September 15th in the calculation for that site. The numerical computation, complex and out of the scope of this document, was done using the open-source package {\tt astroplan} \citep{astroplan}. We performed the calculation over the entire year 2019 and remark that the exact results may slightly differ for different years due to Moon phases. With the aforementioned constraints, we conclude that J0847.0-2336 and J1714.0-2029 are observable from H.E.S.S. and CTA-S for over 600 hours every year, while for MAGIC and CTA-N only J1714.0-2029 is barely visible for 88 hours per year given the aforemention conditions. VERITAS, at slighly higher latitude, can observe both sources only above a zenith distance of $50^{\circ}$. A table containing the observability for the entire sample can be found in Appendix \ref{sec:observability}.

\subsection{Detectability by atmospheric Cherenkov telescopes}

A common approach for estimating the detectability by IACTs is to extrapolate {\it Fermi}-LAT HE spectra \citep[see e.g., ][]{2010PASJ...62.1005I,2019scta.book.....C,2021JCAP...02..048A}. However, this method neglects the information available at longer wavelengths and is not sensitive to possible spectral breaks or curvature affecting the VHE regime. 
In order to reduce such bias, at the cost of model dependency, we designed a method that uses the information contained in the broad-band spectra. First, we produce simulated Monte Carlo (MC) event samples in the range of $100\,\mathrm{GeV}$ to $10\,\mathrm{TeV}$, with a simulated event spectrum that matches that of the extrapolation at $\gamma$-ray energies from the multi-wavelength models described in section \ref{sec:model}.
Then, we analyze these simulated events with standard VHE data analysis tools to evaluate the detectability of each source.

MC simulations are generated using the {\tt ctobssim} tool from the {\tt ctools} package \citep{ctools} on its version 1.7.2. We used CTA's public {\tt prod3b-v2} IRFs \footnote{\url{https://www.cta-observatory.org/science/ctao-performance/}} as a test-bench for VHE detectability. These IRFs are based on the ``Omega Configuration'' of CTA for their north and south hemisphere sites and include the full array that could potentially be deployed at both sites.
The blazars, assumed to be point-like sources, were located at the coordinates available in the 4FGL. 
To simplify the analysis, we do not consider other possible sources in the ROI. The assumed background is extracted from the {\tt prod3b-v2} IRFs. It depends on the longitude and latitude offset, with respect to the center of the field of view, and the energy. It however neglects differences in the background due to the position of the source with respect, for example, to our Galaxy or other sources. The simulated spectrum of the background follows approximately a power-law with spectral index $\alpha \sim 2.57$. 
Consequently, the assumption of the $\gamma$-ray source position has no major implications in the analysis, despite the better angular resolution of IACTs at energies above a few hundreds of GeV. 
Instead, the sole role of the source location is to allow for the most appropriate IRFs to be selected. 
For example, sources with positive declination are matched to the north site IRFs, while south site IRFs are used for sources with negative declination.
The declination is also used to estimate the maximum transit altitude attainable by the sources at either Roque de los Muchachos or Paranal, in order to select the best matching zenith-angle dependent IRF files. 

The energy range was set conservatively from $100\,\mathrm{GeV}$ to $10\,\mathrm{TeV}$. We selected an exposure time of $5\,\mathrm{h}$. On one hand, CTA observing time will be highly demanded and we believe that this is a reasonable exposure that could be requested for individual sources. On the other hand, the sensitivities of CTA-N and CTA-S for just $5\,\mathrm{h}$ in that energy range are in rough approximation similar to the ones from VERITAS, MAGIC and H.E.S.S respectively in $50\,\mathrm{h}$\footnote{See, e.g.,  \protect\url{https://www.cta-observatory.org/science/ctao-performance/}}. The energy threshold of the analysis, expected to be lower for CTA, is set higher in this analysis to match that of existing IACTs. Therefore, these simulations are a decent approximation of the results we could potentially expect with current generation IACTs in $50\,\mathrm{h}$.

Using {\tt ctools}, we performed a simple 3D analysis of the generated samples with the aim of producing spectral energy distributions using {\tt ctlike} and {\tt csspec}. The resulting best-fit spectral data are shown in Figure 4.
Only 4FGL J0847.0-2336 and 4FGL J1714.0-2029 seem to be promising VHE-emitter candidates, with an expected signal strong enough to build a VHE spectrum. The rest are either too faint or too soft to be detectable by the current generation of IACTs in reasonable exposures, and result in no significant detections. The two detectable sources have, according to the proposed emission model, very low magnetization ($B\sim10^{-2}\,\mathrm{G}$) of the plasma, strong host emission (compared to the synchrotron radiation) and no IR torus.

\section{Discussion and conclusions}\label{sec:discussion}

We find that a simple one-zone SSC model reasonably explains the observed emission in X- and $\gamma$-rays for both the selected HSPs and EHSPs. The IR and optical observations can be correctly reproduced using a combination of synchrotron emission and a black body spectrum with effective temperatures compatible with starlight emission from the host, component that is sometimes not detectable in HSPs but seem to be present in many EHSPs. The sharp fall towards low frequencies of that peak, together with effective temperatures that are compatible with a population of predominantly old K-M stars, points towards passively evolving elliptical host galaxies. These galaxies tend to be characterized by low star formation rates and low amounts of dust and gas. 

For J0132.7-0804, J0350.4-5144, J0606.5-4730, J1934.3-2419, J1944.4-4523, J2001.9-5737, this single temperature host model is insufficient to reproduce the observed data. For them, we included an additional torus-like component in the modeling, and evaluated the contribution of external Compton scattering to the high energy photon spectra. Adding a torus contributes also to improve the reproduction of very curved $\gamma$-ray spectral, observed for some of the sources. For high-power blazars (e.g. FSRQs), this component often outshines the SSC emission and usually results in very large $\gamma$-ray fluxes and large Compton dominance values \citep[CD, see, e.g.,][]{finke13,paliya17}. This is not the only solution since a more complex system such as multiple emitting regions or a richer host emission spectra could also reproduce the emission, but we propose the more simple answer.

For a few sources, for example J0212.2-0219, J0953.4-7659 and J2246.7-5207, we observe a large dispersion of the measured X-ray fluxes, possibly indicating that these sources are indeed variable. In contrast, the $\gamma$-ray data do not exhibit such large scale variability, either due to lack of sensitivity of {\em Fermi}-LAT detecting short-term flux changes, or perhaps, because of wavelength-dependent variability. The latter would be consistent with a more complicated emission scenario, for example from a structured jet. We have neglected these effects in this work.

Only for J0350.4-5144, J1944-4523 and 4FGLJ2001.9-5737, with photometric redshifts of $z=0.32$, $z=0.21$ and $z=0.26$ respectively, the external Compton emission in $\gamma$-rays seems to be brighter than the SSC component. Even then, and as it occurs with most of the other sources, CD$<1$.
Following \cite{bonnoli2015}, we conclude that the observed spectra and the small CD measured are a result of a low magnetization of the jet at the position where the $\gamma$-rays are produced, together with a lack of strong external photon fields which could serve as seed photons for the external Compton process, 
in line for that of other EHSPs such as 1ES0229+200. 
On the other hand, \cite{bonnoli2015} observed that large $\gamma_{\rm min}$ values are usually required to explain the hard intrinsic spectra at TeV energies, unless lepto-hadronic models or external Compton scattering on the CMB is invoked to explain it. 
This observation was further supported by \cite{2018MNRAS.477.4257C} using a sample of known TeV emitters.
Nonetheless, our sample of potential HSP and EHSP do not show exceptionally hard spectra in sub-TeV energies. 
However, high values of $\gamma_{\rm min}$ are still favored by the hard spectra exhibited by some sources below $\sim 1\,\mathrm{GeV}$.

VHE observations with existing IACT facilities or the future CTA are probably needed to shed more light on this open questions. In the one-zone SSC scenario, VHE observations can also potentially constrain the synchrotron peak location indirectly. 
Magnetic fields are regarded as a key element in the current picture of particle acceleration in AGN jets. Magnetic reconnection provides a way of converting the initially Poynting flux-dominated jet energy into kinetic energy of the particles embedded in it \citep{2007MNRAS.380...51K}, a mechanism that operates efficiently until equipartition is reached ($U_e \approx U_B$). For BLLs, however, simple one-zone emission models are often in conflict with this prediction, i.e. $U_B \ll U_e$. 
For the sources we modelled with a pure, one-zone, SSC model, we find that most have values of $U_B / U_e$ ranging from $\sim 10^{-2}$ to $\sim 10^2$, independently of whether they are HSPs or EHSPs.
Moreover, both J0847.0-2336 and J1714.0-2029, the most promising VHE emitter candidates, have the smallest ratio of magnetic to kinetic energy density as shown in Figure \ref{fig:equipartition}. 
In this line, \cite{2016MNRAS.456.2374T}, following previous work from \cite{1998ApJ...509..608T}, proposed an analytical estimate of the ratio between the rest-frame $U_B$ and $U_e$ assuming one-zone SSC emission with typical BL Lac parameters. 
They observed that this ratio depended on the frequency of the maximum of the synchrotron and inverse Compton components and could reach potentially low values for sources whose emission peaks at high frequencies. \cite{2018MNRAS.477.4257C} gave further evidence of this using a sample of extreme blazars with significant TeV emission. They found values as large as $U_e/U_B \sim 10^{6}$ for some objects. This result is often used to support the idea that structured jets with stronger overall magnetic fields \citep{2008MNRAS.385L..98T} may be needed to successfully produce emission at VHE energies, as opposed to the classical one-zone scenario. 
We note however that the estimated energy balance of the blazars is model-dependent, and for some sources (e.g. J0212.2-0219, J0958.1-6753, J1132.2-4736) the emission in VHE is not well constrained by the available data. For such cases, we favored conservative model predictions with smaller maximum Lorentz factor $\gamma_{\rm max}$ and lower emitted flux in the VHE band. As a result, the measured ratio of magnetic to kinetic energy density might be biased towards higher values, closer to equipartition. Further measurements of the emitted flux from these blazars in hard X-rays and VHE $\gamma$-rays would be needed to confirm the energy density predictions from this study.

In conclusion, extreme blazars are one of the most interesting members of the extragalactic $\gamma$-ray source class, lying at the very end of the blazar sequence. Even though they are typically referred as being good very-high-energy $\gamma$-ray candidates, the population of these sources with detected $\gamma$-ray fluxes remains small because of their low fluxes and their steady emission compared to most other blazars.

In this paper we have conducted a search for new potential extreme blazars among poorly characterized sources in {\em Fermi}-LAT's BCU list. 
The comparison of the location of these BCUs with position of known infrared sources with properties compatible to those of HSP blazars resulted in 23 candidates, including J0733.4+5152, detected in VHE by MAGIC \citep{j0733} and classified as an EHSP in that work. For the other 22,
the analysis of multi-wavelength observations carried from multiple surveys plus a re-analysis of {\em Fermi}-LAT data focusing on the highest available $\gamma$-ray energies allowed us to identify 17 new potential EHSP and five less extreme (HSP) blazars. 
The emission is modelled and interpreted as SSC radiation plus, in some cases, external Compton from a relatively hot, $2\times10^2 \lesssim \mathrm{T_{DT} (K)} \lesssim 1.2\times10^3$, dusty torus component in six of these sources. The resulting models not only reproduce the observed spectra, but also allowed us to: i) classify the sources according to the frequency of the synchrotron peak maximum; ii) identify two potentially detectable extreme blazars in the VHE band (J0847.0-2336 and J1714.0-2029);  iii) evaluate the balance between magnetic and kinetic energy density for each of them. As a result, we found that the most promising VHE candidates are the ones deviating the most from equipartition. 

\section*{Acknowledgements}
The \textit{Fermi}-LAT Collaboration acknowledges generous ongoing support
from a number of agencies and institutes that have supported both the
development and the operation of the LAT as well as scientific data analysis.
These include the National Aeronautics and Space Administration and the
Department of Energy in the United States, the Commissariat \`a l'Energie Atomique
and the Centre National de la Recherche Scientifique / Institut National de Physique
Nucl\'eaire et de Physique des Particules in France, the Agenzia Spaziale Italiana
and the Istituto Nazionale di Fisica Nucleare in Italy, the Ministry of Education,
Culture, Sports, Science and Technology (MEXT), High Energy Accelerator Research
Organization (KEK) and Japan Aerospace Exploration Agency (JAXA) in Japan, and
the K.~A.~Wallenberg Foundation, the Swedish Research Council and the
Swedish National Space Board in Sweden.

Additional support for science analysis during the operations phase is gratefully
acknowledged from the Istituto Nazionale di Astrofisica in Italy and the Centre
National d'\'Etudes Spatiales in France. This work performed in part under DOE
Contract DE-AC02-76SF00515.

This research made use of ctools, a community-developed analysis package for Imaging Air Cherenkov Telescope data. ctools is based on GammaLib, a community-developed toolbox for the scientific analysis of astronomical gamma-ray data.

M.N. is thankful for the support of the Deutsches Elektronen-Synchrotron (DESY), the Young Investigator Program of the Helmholtz Association in Germany, and the Instituto de Astrofisica de Canarias (IAC, contract PS-2019-073 Astroparticulas-CTA 2019), during the duration of this project. 
A.D. is thankful for the support of the Ram{\'o}n y Cajal program from the Spanish MINECO.

\section*{Data availability}

The data underlying this article will be shared on reasonable request to the corresponding author.


\newpage

\begin{appendix}

\section{Visibility of the selected blazars from the different IACT sites}\label{sec:observability}

Table \ref{tab:observ} summarizes the resulting number of hours applying these criteria to the sources as seen from the different sites. 

\begin{table}
\caption{Observability, in number of hours per year during astronomical nights with Moon illumination less than $50\%$ (or below horizon) and the source above $40\,\mathrm{deg}$ elevation, of the different sources from major IACT observatories. Note that Roque de los Muchachos (ORM) hosts both MAGIC and the future CTA-N, therefore they have the same observability. VERITAS is referred as VTS in this table for compactness. The calculation of the observability for the VERITAS site takes into account the shutdown in Summer due to the Monsoon.}
\label{tab:observ}
\begin{tabular}{lcrrrr} \toprule
\toprule
4FGL Name  & MAGIC & CTA-N &  VTS &  HESS &  CTA-S \\
\midrule
J0132.7$-$0804 &    396 &    396 &      280 &   528 &    532 \\
J0212.2$-$0219 &    474 &    474 &      374 &   490 &    496 \\
J0350.4$-$5144 &      0 &      0 &        0 &   568 &    595 \\
J0515.5$-$0125 &    504 &    504 &      472 &   491 &    488 \\
J0526.7$-$1519 &    298 &    298 &      216 &   560 &    562 \\
J0529.1+0935 &    600 &    600 &      573 &   407 &    400 \\
J0557.3$-$0615 &    455 &    455 &      417 &   528 &    522 \\
J0606.5$-$4730 &      0 &      0 &        0 &   600 &    616 \\
J0647.0$-$5138 &      0 &      0 &        0 &   598 &    616 \\
J0847.0$-$2336 &      0 &      0 &        0 &   622 &    628 \\
J0953.4$-$7659 &      0 &      0 &        0 &     0 &      0 \\
J0958.1$-$6753 &      0 &      0 &        0 &   430 &    480 \\
J1132.2$-$4736 &      0 &      0 &        0 &   656 &    677 \\
J1447.0$-$2657 &      0 &      0 &        0 &   664 &    672 \\
J1714.0$-$2029 &     88 &     88 &        0 &   639 &    642 \\
J1824.5+4311 &    613 &    613 &      344 &     0 &      0 \\
J1934.3$-$2419 &      0 &      0 &        0 &   635 &    637 \\
J1944.4$-$4523 &      0 &      0 &        0 &   645 &    656 \\
J2001.9$-$5737 &      0 &      0 &        0 &   587 &    612 \\
J2142.4+3659 &    610 &    610 &      316 &     0 &      0 \\
J2246.7$-$5207 &      0 &      0 &        0 &   600 &    620 \\
J2251.7$-$3208 &      0 &      0 &        0 &   620 &    632 \\
\bottomrule
\end{tabular}
\end{table}

\section{Model results using $\gamma_{\rm min}=1$}\label{sec:gammamin1}

Besides the preferred model described in section \ref{sec:model}, for which we assumed a minimum Lorentz factor $\gamma_{\rm min} = 10^3$, we also explored the possibility of having a lower Lorentz factor $\gamma_{\rm min} = 1$. The resulting broadband SEDs provide a reasonable reconstruction of the observed broadband spectra, as it can be seen in Figure \ref{fig:mw_seds_gammamin1}. For most sources and wavelengths, the model predictions are comparable to the one from models with $\gamma_{\rm min} = 10^3$. 
The exceptions to this are the lowest energies detectable by {\em Fermi}-LAT, below $\sim 1\,\mathrm{GeV}$. At such energies, several sources, including J0515.5-0125, J0526.7-1519, J1714.0-2029, J1934.3-2419, J1944.4-4523, J2246.7-5207, and J2251.7-3208, exhibit very hard and curved spectra.

We note that in order to directly constrain $\gamma_\mathrm{min}$, observations in the far and mid-IR, together with measurements with medium and low energy  gamma-rays with enough sensitivity would be needed. However,  $\gamma_\mathrm{min}$ has other effects that are sensitive to with the described modeling approach. 
In a few cases it required an unreasonably hard electron spectral index $p_1$, and even then, the reconstructed $\gamma$-ray spectrum is unable to convincingly reproduce the observations. 
Following \cite{tavecchio2010} and \cite{kaufmann2011}, we conclude that this result supports the idea of a higher value of the minimum Lorentz factor needed to reproduce the spectrum of some HSP and EHSP blazars. 

\begin{figure}
\centering
\includegraphics[width=0.48\textwidth]{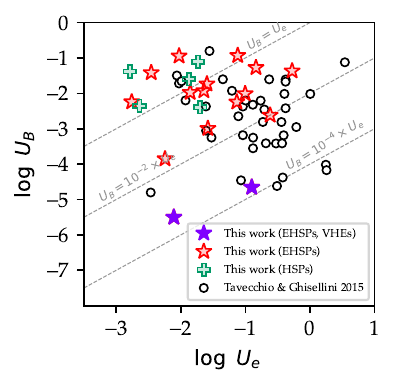}
\caption{
Magnetic energy density as a function of kinetic energy density derived from our sample (red and purple stars for sources classified as EHSP and green crosses for HSP classified objects) and the one for BLLs presented in \protect\cite{2016MNRAS.456.2374T} (open black circles). The modeling was produced assuming $\gamma_{\rm min}=1$.
Dashed gray lines show the $U_B = U_e$ (equipartition), $U_B=10^{-2} \times U_e$ and $U_B=10^{-4} \times U_e$ cases. 
According to the model, our sources have on average slightly more magnetized jets for a given kinetic energy that the sample of BLLs presented in \protect\cite{2016MNRAS.456.2374T}, and closer in general to equipartition (except for the two VHE candidates).
}
\label{fig:equipartition_gammamin1}
\end{figure}

Finally, we estimated again how close the different blazar models are from equipartition, representing the magnetic energy density as a function of kinetic energy density in Figure \ref{fig:equipartition_gammamin1}. Compared to the model with $\gamma_{\rm min}=10^3$, the EHSP jets are less magnetized and more kinetically dominated in models with $\gamma_{\rm min}=1$, but we note that the conclusions about EHSP blazars with VHE emission being further away from equipartition remain unchanged.

\begin{figure*}
\centering
\includegraphics[width=0.325\linewidth]{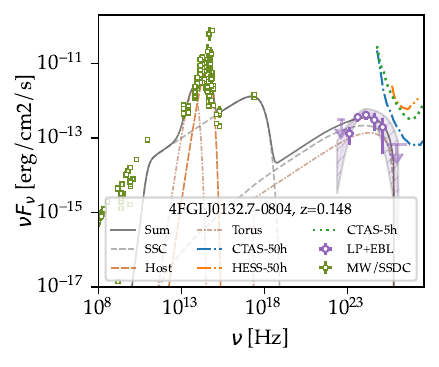}
\includegraphics[width=0.325\linewidth]{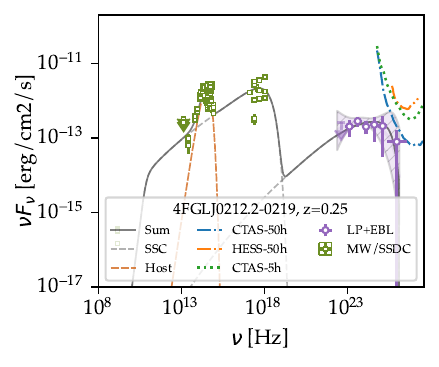}
\includegraphics[width=0.325\linewidth]{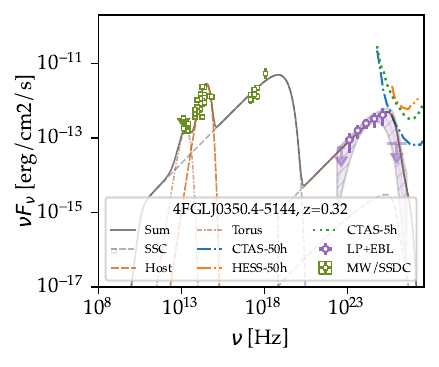}
\includegraphics[width=0.325\linewidth]{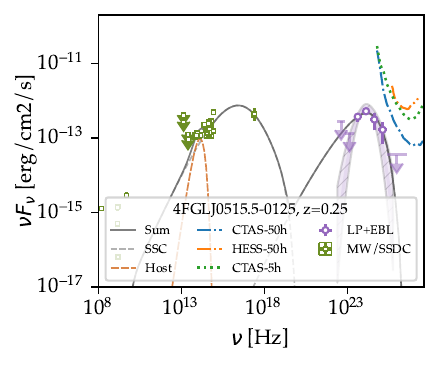}
\includegraphics[width=0.325\linewidth]{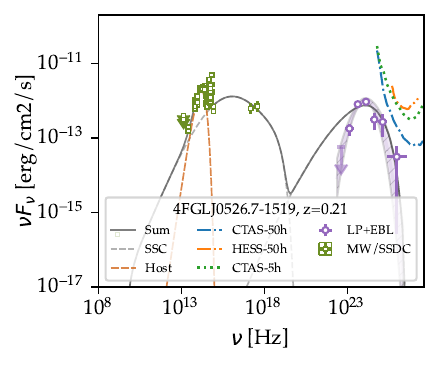}
\includegraphics[width=0.325\linewidth]{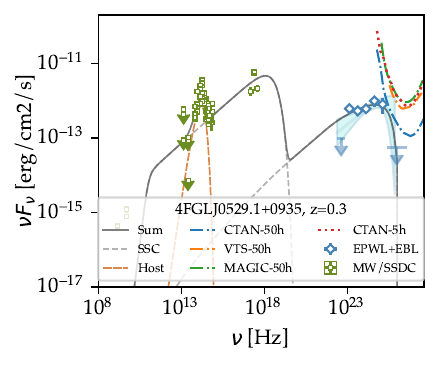}
\includegraphics[width=0.325\linewidth]{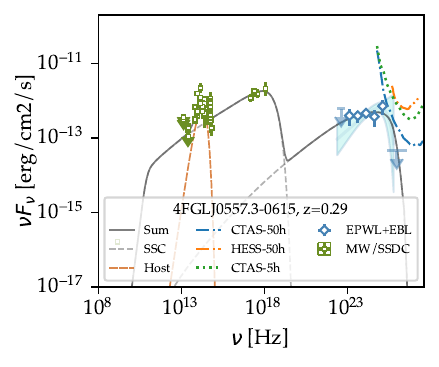}
\includegraphics[width=0.325\linewidth]{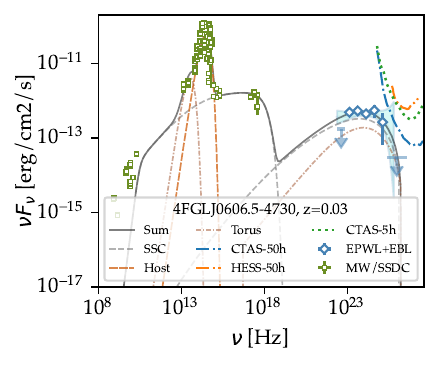}
\includegraphics[width=0.325\linewidth]{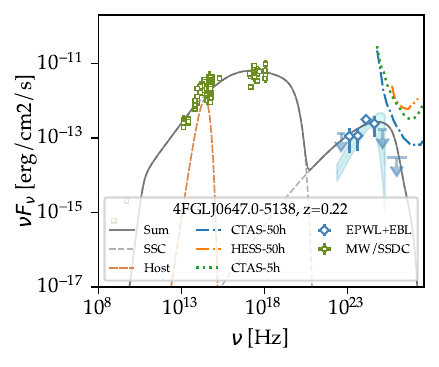}
\includegraphics[width=0.325\linewidth]{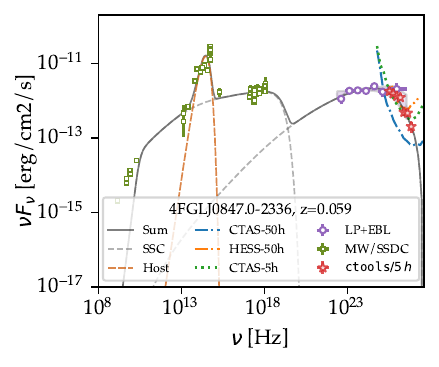}
\includegraphics[width=0.325\linewidth]{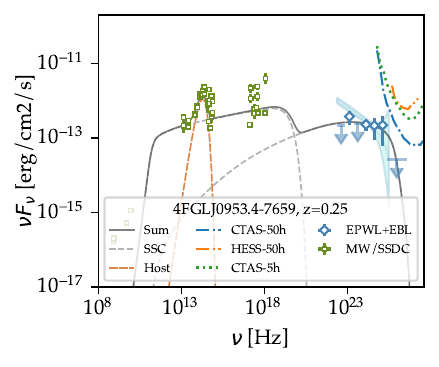}
\includegraphics[width=0.325\linewidth]{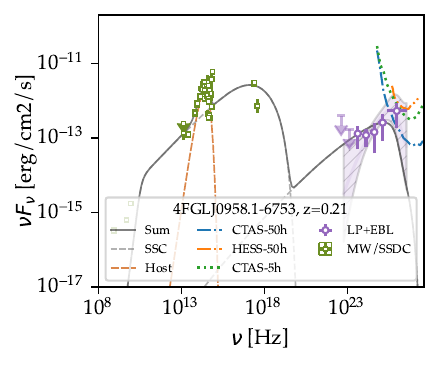}
\includegraphics[width=0.325\linewidth]{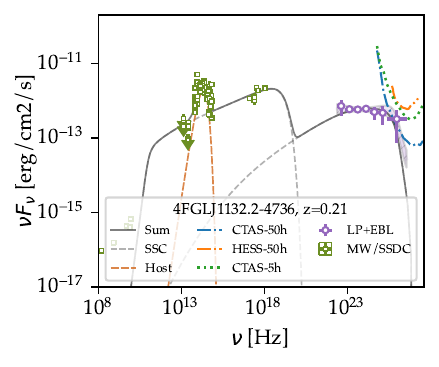}
\includegraphics[width=0.325\linewidth]{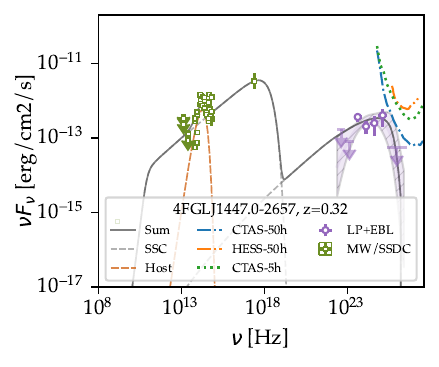}
\includegraphics[width=0.325\linewidth]{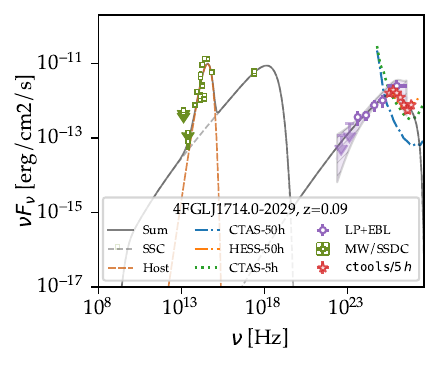}
\label{fig:mw_seds_gammamin1_0}
\end{figure*}

\begin{figure*}
\ContinuedFloat
\centering
\includegraphics[width=0.325\linewidth]{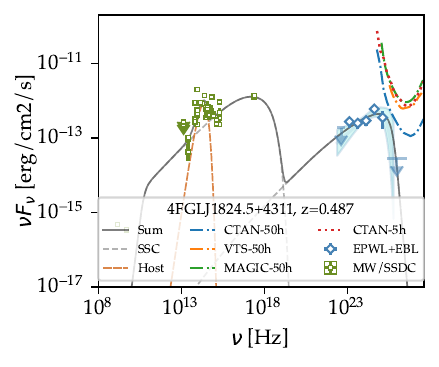}
\includegraphics[width=0.325\linewidth]{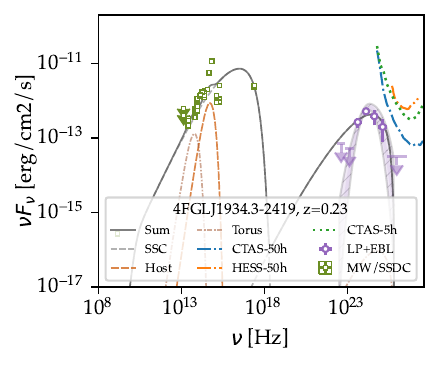}
\includegraphics[width=0.325\linewidth]{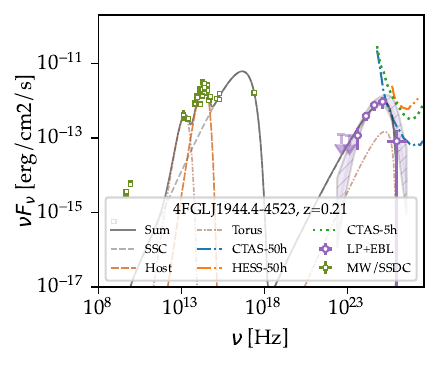}
\includegraphics[width=0.325\linewidth]{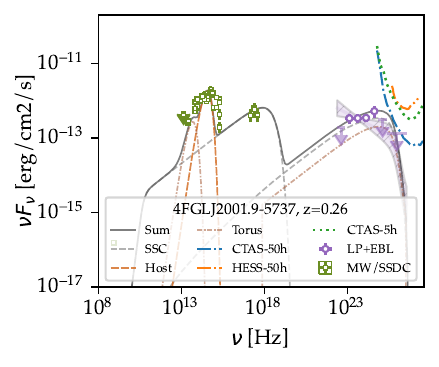}
\includegraphics[width=0.325\linewidth]{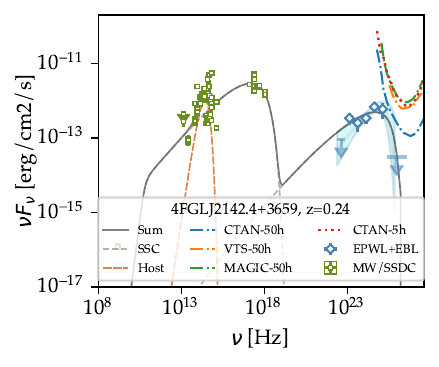}
\includegraphics[width=0.325\linewidth]{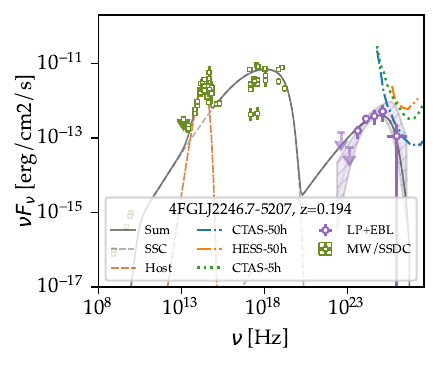}
\includegraphics[width=0.325\linewidth]{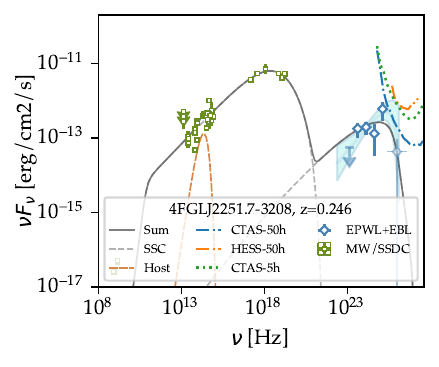}
\caption{Multi-wavelength emission of the selected sources, including archival measurements, the re-analysis of HE $\gamma$-ray data from {\em Fermi}-LAT and `toy' Monte Carlo based SED reconstructed with {\tt ctools} from the Prod-3b CTA simulations, described in section \ref{sec:vhe_prospects}. In all cases, the one-zone leptonic emission model, with fixed minimum Lorentz factor $\gamma_{\rm min} = 1$, plus thermal components (host emission and possibly a dusty torus)} has been fitted to the measured points (excluding {\tt ctools} points). 
\label{fig:mw_seds_gammamin1}
\end{figure*}

The corresponding model parameters and luminosity of the different components are shown in Tables \ref{tab:jet_parameters_gammamin1} and \ref{tab:model_parameters_gammamin1}, following the same format as for the models with $\gamma_{\rm min} = 10^3$.

\begin{landscape}
\begin{table}
\caption{Main jet model best-fit parameters, including electron density $\mathrm{N}$, 
magnetic field strength $\mathrm{B}$, 
electron indices before and after the break $p_1$ and $p_2$, 
maximum Lorentz factors $\gamma_\mathrm{max}$ and position of the spectral break $\gamma_\mathrm{br}$, 
jet angle with respect to the line of sight $\theta$, 
Doppler boosting $\delta$, 
estimated position of the synchrotron peak $\nu_\mathrm{SP}$, 
favored LAT model to reproduce the high energy data: Log Parabola (LP) or Power Law with exponential cut-off (EPWL),  
Compton dominance (CD) and the classification as either HSP or EHSP. 
Fixed values: radius of the emitting region $\mathrm{R}=1.0\times 10^{16}\, \mathrm{cm}$, position with respect to the central object $\mathrm{R_H}=2.0\times 10^{18}\, \mathrm{cm}$, minimum Lorentz factor $\gamma_{\rm min}=1$, bulk Lorentz factor $\Gamma=20$.} [$\dagger$] denotes unconstrained best-fit parameters.
\centering
\label{tab:jet_parameters_gammamin1}
\begin{tabular}{lrrrrrrrrrrrrr}
\toprule
4FGL Name & $\mathrm{N}$ & $\mathrm{B}$ & $p_1$ & $p_2$ & $\gamma_\mathrm{br}$ & $\gamma_\mathrm{max}$ & $\theta$ & $\delta$ & $\log{\mathrm{\nu_{SP}}}$ & LAT model & CD & Class\\
& & & & & $[\times 10^{4}]$ & $[\times 10^{6}]$ & & & & & \\
& $\mathrm{(cm^{-3})}$ & $\mathrm{(G)}$ & & & & & $\mathrm{(deg})$ & & $\mathrm{[Hz]}$ & & \\
\midrule
J0132.7$-$0804 &   $(7.4\pm0.7)\times 10^{4}$ &     $1.16\pm0.05$ &   $2.410\pm0.009$ &     $2.4\pm0.8$ &     $(6\pm6)\times 10^{2}$ &     $0.180\pm0.004$ &     $7.8\pm1.6$ & $4.8$ &     $17.2$ &        LP &  $0.27$ &  EHSP \\
J0212.2$-$0219 &   $(2.8\pm2.5)\times 10^{3}$ &     $0.68\pm0.06$ &     $1.90\pm0.08$ &   $2.38\pm0.15$ &           $0.03 [\dagger]$ &     $0.296\pm0.008$ &     $5.3\pm0.4$ & $9.1$ &     $17.7$ &        LP &  $0.13$ &  EHSP \\
J0350.4$-$5144 &   $3\times 10^{1} [\dagger]$ &     $0.52\pm0.26$ &     $2.09\pm0.08$ &     $2.2\pm1.4$ &                  $16\pm11$ &         $0.6\pm0.4$ &     $1.1\pm0.6$ &  $35$ &     $18.8$ &        LP &  $0.13$ &  EHSP \\
J0515.5$-$0125 &                $2 [\dagger]$ &       $1.4\pm0.6$ &     $1.06\pm0.05$ &     $5.9\pm2.0$ &                $6.3\pm1.7$ &     $0.695\pm0.017$ &     $8.9\pm1.4$ & $3.8$ &     $16.4$ &        LP &  $0.62$ &   HSP \\
J0526.7$-$1519 &              $0.3 [\dagger]$ &     $0.79\pm0.04$ &     $0.75\pm0.05$ &     $4.2\pm0.4$ &                $1.5\pm0.6$ &     $0.559\pm0.008$ &     $6.5\pm0.8$ & $6.4$ &     $16.1$ &        LP &  $0.57$ &   HSP \\
J0529.1$+$0935 &   $(2.1\pm0.8)\times 10^{4}$ &       $1.7\pm0.9$ &   $2.207\pm0.031$ &   $2.21\pm0.06$ &                $0.6\pm0.4$ &       $0.36\pm0.16$ &     $7.3\pm1.3$ & $5.4$ &     $18.1$ &      EPWL &  $0.17$ &  EHSP \\
J0557.3$-$0615 & $1.4\times 10^{4} [\dagger]$ &   $0.378\pm0.014$ &     $2.06\pm0.08$ &   $2.45\pm0.14$ &            $0.053\pm0.025$ &     $0.491\pm0.005$ &     $4.8\pm1.8$ &  $10$ &     $17.9$ &      EPWL &  $0.27$ &  EHSP \\
J0606.5$-$4730 & $2.2\times 10^{3} [\dagger]$ &     $0.32\pm0.04$ &     $1.94\pm0.09$ &   $2.99\pm0.28$ &            $0.5 [\dagger]$ &     $0.4 [\dagger]$ &     $7.1\pm1.9$ & $5.6$ &     $16.7$ &      EPWL &  $0.32$ &   HSP \\
J0647.0$-$5138 & $0.6\times 10^{2} [\dagger]$ &     $0.38\pm0.28$ &     $1.75\pm0.05$ &     $3.3\pm0.5$ &            $1.9 [\dagger]$ &     $1.3 [\dagger]$ &   $1.93\pm0.35$ &  $28$ &     $17.2$ &      EPWL &  $0.09$ &  EHSP \\
J0733.4$+$5152 &   $(7.0\pm1.9)\times 10^{3}$ &     $0.36\pm0.23$ &   $2.259\pm0.030$ &   $3.41\pm0.23$ &                   $15\pm7$ &       $1.25\pm0.10$ &     $5.1\pm1.9$ & $9.6$ &     $17.9$ &      EPWL & $0.081$ &  EHSP \\
J0847.0$-$2336 & $(2.27\pm0.29)\times 10^{4}$ &  $0.02 [\dagger]$ &   $2.061\pm0.015$ &   $2.89\pm0.08$ &            $0.5 [\dagger]$ &         $2.4\pm1.3$ &   $3.14\pm0.29$ &  $18$ &     $17.9$ &        LP &   $1.3$ &  EHSP \\
J0953.4$-$7659 &   $(1.9\pm1.1)\times 10^{5}$ &       $1.0\pm0.8$ &     $2.18\pm0.14$ &   $2.83\pm0.27$ &          $0.015 [\dagger]$ &       $1.21\pm0.04$ &     $6.8\pm0.5$ & $6.0$ &     $18.5$ &      EPWL &  $0.43$ &  EHSP \\
J0958.1$-$6753 & $1.0\times 10^{3} [\dagger]$ &   $0.060\pm0.033$ &     $2.08\pm0.06$ &     $8.2\pm1.1$ &     $(9\pm8)\times 10^{1}$ &         $0.9\pm0.5$ & $0.0 [\dagger]$ &  $40$ &     $17.1$ &        LP &   $0.1$ &  EHSP \\
J1132.2$-$4736 & $(5.80\pm0.24)\times 10^{3}$ &   $0.495\pm0.032$ &   $1.553\pm0.009$ & $2.621\pm0.018$ &          $0.0100\pm0.0016$ &     $0.971\pm0.033$ &   $5.50\pm0.11$ & $8.5$ &     $18.4$ &        LP &   $0.3$ &  EHSP \\
J1447.0$-$2657 & $0.5\times 10^{4} [\dagger]$ &       $0.5\pm0.4$ &     $2.10\pm0.06$ &   $2.26\pm0.13$ &          $0.016 [\dagger]$ &       $0.25\pm0.05$ &     $4.0\pm1.6$ &  $14$ &     $17.7$ &        LP & $0.097$ &  EHSP \\
J1714.0$-$2029 &                $7 [\dagger]$ & $0.009 [\dagger]$ &     $1.26\pm0.07$ &   $1.94\pm0.15$ &            $0.079\pm0.031$ &         $1.9\pm1.4$ &     $1.6\pm1.2$ &  $31$ &     $18.2$ &        LP &  $0.29$ &  EHSP \\
J1824.5$+$4311 &   $(4.2\pm1.6)\times 10^{3}$ &     $0.16\pm0.10$ &   $2.072\pm0.025$ &     $4.9\pm1.9$ &                  $33\pm24$ &     $0.494\pm0.024$ &   $2.51\pm0.29$ &  $23$ &     $17.2$ &      EPWL &  $0.35$ &  EHSP \\
J1934.3$-$2419 &                $0.47\pm0.08$ &     $1.02\pm0.15$ &   $1.106\pm0.012$ &     $4.1\pm0.5$ &              $2.37\pm0.26$ &     $0.054\pm0.004$ &   $3.63\pm0.35$ &  $15$ &     $16.5$ &        LP & $0.059$ &   HSP \\
J1944.4$-$4523 &             $0.06 [\dagger]$ &     $0.33\pm0.06$ &     $1.00\pm0.05$ &     $5.2\pm1.2$ &                  $30\pm22$ & $0.07553\pm0.00006$ &     $3.5\pm1.3$ &  $16$ &     $16.7$ &        LP &  $0.16$ &   HSP \\
J2001.9$-$5737 & $(9.34\pm0.24)\times 10^{4}$ &   $0.246\pm0.034$ &   $2.311\pm0.005$ & $2.742\pm0.025$ &            $160.01\pm0.26$ &       $0.73\pm0.05$ &   $6.35\pm0.12$ & $6.8$ &     $17.9$ &        LP &  $0.84$ &  EHSP \\
J2142.4$+$3659 & $(1.60\pm0.20)\times 10^{3}$ &       $0.5\pm0.4$ &   $2.000\pm0.013$ &     $4.0\pm1.3$ &               $12.6\pm2.5$ &       $0.24\pm0.07$ &     $4.4\pm1.4$ &  $12$ &     $17.1$ &      EPWL &  $0.17$ &  EHSP \\
J2246.7$-$5207 &                 $24.5\pm3.2$ &   $0.981\pm0.025$ & $1.6053\pm0.0017$ & $3.412\pm0.019$ &                    $7\pm4$ &     $0.739\pm0.017$ & $5.268\pm0.018$ & $9.1$ &     $18.1$ &        LP & $0.059$ &  EHSP \\
J2251.7$-$3208 &   $(1.4\pm1.1)\times 10^{3}$ &     $1.69\pm0.20$ &     $2.07\pm0.13$ &    $11.0\pm1.3$ & $(2.4\pm0.8)\times 10^{2}$ &       $2.33\pm0.12$ &     $5.8\pm0.5$ & $7.8$ &     $18.4$ &      EPWL &  $0.16$ &  EHSP \\
\bottomrule
\end{tabular}
\end{table}
\end{landscape}

\begin{landscape}
\begin{table}
\centering
\caption{Energy budget, showing the effective temperature $\mathrm{T_{eff,host}}$ of the black body that we added to simulate host galaxy emission, the effective temperature of the dusty torus $\mathrm{T_{DT}}$, 
its distance $\mathrm{R_{DT}}$ from the core and its opacity $\tau_\mathrm{DT}$, 
the integrated host luminosity $\mathrm{L_{host}}$ and the luminosity carried by the jet for the non-thermal low energy and total radiative components $\mathrm{L_{sync}}$ and $\mathrm{L_{rad}}$, the electrons $\mathrm{L_{kin}}$, 
the Poynting luminosity due to the magnetic field} $\mathrm{L_B}$, 
the total jet luminosity $\mathrm{L_{tot}}$ and the energy density ratio of the magnetic field to that of the electron distribution $\mathrm{U_B/U_e}$. 
$[\dagger]$ denotes unconstrained best-fit parameters.
\label{tab:model_parameters_gammamin1}
\begin{tabular}{lrrrrrrrrrrr}
\toprule
4FGL Name  & $\mathrm{R_{DT}}$ & $\mathrm{T_{DT}}$ & $\tau_\mathrm{DT}$ & $\mathrm{T_{host}}$ & $\mathrm{L_{host}}$ & $\mathrm{L_{sync}}$ & $\mathrm{L_{rad}}$ & $\mathrm{L_{B}}$ & $\mathrm{L_{kin}}$ & $\mathrm{L_{tot}}$ & $\mathrm{U_B/U_e}$ \\
& $[\times 10^{18}]$ & $[\times 10^{2}]$ & & $[\times 10^{3}]$ & $[\times 10^{44}]$ & $[\times 10^{42}]$ & $[\times 10^{42}]$ & $[\times 10^{42}]$ & $[\times 10^{44}]$ & $[\times 10^{44}]$ & \\
& $\mathrm{(cm)}$ & $\mathrm{(K)}$ & & $\mathrm{(K)}$ & $\mathrm{(erg/s)}$ & $\mathrm{(erg/s)}$ & $\mathrm{(erg/s)}$ & $\mathrm{(erg/s)}$ & $\mathrm{(erg/s)}$ & $\mathrm{(erg/s)}$ & \\
\midrule
J0132.7$-$0804 &                $3.3\pm0.7$ & $7.53\pm0.19$ &    $0.25\pm0.05$ &   $5.2\pm0.5$ &     $6.6\pm1.4$ &              $94$ & $2.0\times10^{2}$ & $1.3\times10^{2}$ &              $5.5$ &     $8.8$ &             $0.37$ \\
J0212.2$-$0219 &                          - &             - &                - &   $5.3\pm1.5$ &     $5.7\pm2.7$ &              $37$ &              $69$ &              $43$ &             $0.96$ &     $2.1$ &             $0.72$ \\
J0350.4$-$5144 &              $3 [\dagger]$ &   $3.1\pm1.0$ & $0.06 [\dagger]$ &   $5.2\pm0.8$ &        $14\pm5$ &            $0.56$ &              $41$ &            $0.65$ & $5.9\times10^{-3}$ &    $0.42$ &               $69$ \\
J0515.5$-$0125 &                          - &             - &                - & $2.07\pm0.11$ &   $0.27\pm0.24$ & $4.1\times10^{2}$ & $2.9\times10^{2}$ & $6.4\times10^{2}$ &              $0.7$ &      $10$ &              $4.2$ \\
J0526.7$-$1519 &                          - &             - &                - &   $2.5\pm0.8$ &     $3.4\pm1.7$ &              $69$ &              $95$ & $1.0\times10^{2}$ &              $0.5$ &     $2.5$ &              $1.9$ \\
J0529.1$+$0935 &                          - &             - &                - &   $2.5\pm0.6$ &     $3.1\pm1.3$ & $8.7\times10^{2}$ & $4.4\times10^{2}$ & $1.0\times10^{3}$ &              $2.9$ &      $18$ &              $1.5$ \\
J0557.3$-$0615 &                          - &             - &                - &   $2.9\pm0.9$ &     $2.6\pm1.4$ &              $26$ &              $21$ &              $36$ &              $2.8$ &     $3.3$ &            $0.077$ \\
J0606.5$-$4730 &              $5 [\dagger]$ &   $4.4\pm2.6$ & $0.07 [\dagger]$ &   $3.9\pm0.5$ &     $1.8\pm1.2$ &             $3.4$ &              $15$ &             $4.3$ &             $0.75$ &    $0.95$ &              $0.2$ \\
J0647.0$-$5138 &                          - &             - &                - &   $4.0\pm2.1$ &     $2.2\pm1.9$ &             $1.6$ &              $22$ &             $1.7$ &            $0.065$ &     $0.3$ &              $3.3$ \\
J0733.4$+$5152 &                          - &             - &                - & $4.42\pm0.12$ &     $2.0\pm1.5$ &             $5.3$ &              $19$ &             $5.8$ &             $0.79$ &     $1.0$ &             $0.24$ \\
J0847.0$-$2336 &                          - &             - &                - &   $3.8\pm0.7$ &     $1.9\pm1.4$ &            $0.15$ &           $0.083$ &            $0.34$ &              $4.7$ &     $4.7$ & $1.7\times10^{-4}$ \\
J0953.4$-$7659 &                          - &             - &                - &   $3.0\pm0.9$ &     $3.1\pm1.5$ & $1.1\times10^{2}$ & $1.6\times10^{2}$ & $1.6\times10^{2}$ &               $20$ &      $23$ &            $0.081$ \\
J0958.1$-$6753 &                          - &             - &                - & $4.05\pm0.06$ &     $3.7\pm2.1$ &           $0.097$ &            $0.53$ &            $0.11$ &             $0.21$ &    $0.22$ &            $0.025$ \\
J1132.2$-$4736 &                          - &             - &                - &   $3.1\pm0.6$ &     $3.6\pm0.7$ &              $41$ &              $37$ &              $58$ &              $3.7$ &     $4.7$ &            $0.098$ \\
J1447.0$-$2657 &                          - &             - &                - &   $3.0\pm1.0$ &     $2.8\pm2.5$ &              $19$ &              $45$ &              $22$ &             $0.86$ &     $1.5$ &             $0.52$ \\
J1714.0$-$2029 &                          - &             - &                - &   $5.1\pm0.9$ &     $2.9\pm1.0$ &           $0.093$ &           $0.012$ &            $0.13$ &             $0.29$ &     $0.3$ & $4.0\times10^{-4}$ \\
J1824.5$+$4311 &                          - &             - &                - &   $4.0\pm1.2$ &        $11\pm5$ &             $3.1$ &             $3.8$ &             $4.2$ &             $0.99$ &     $1.1$ &            $0.039$ \\
J1934.3$-$2419 & $(2.0\pm0.8)\times 10^{2}$ & $9 [\dagger]$ &    $0.15\pm0.10$ &       $8\pm6$ &     $2.0\pm1.1$ &              $10$ & $1.5\times10^{2}$ &              $11$ &            $0.061$ &     $1.7$ &               $25$ \\
J1944.4$-$4523 &                  $26\pm15$ &   $2.4\pm0.8$ &    $0.14\pm0.11$ &   $3.4\pm0.7$ &     $5.4\pm2.0$ &             $4.5$ &              $17$ &             $5.2$ &            $0.086$ &    $0.31$ &              $1.9$ \\
J2001.9$-$5737 &    $2 [\dagger]$ &   $7.5\pm0.6$ &  $0.148\pm0.007$ &   $5.7\pm0.5$ &     $5.6\pm0.5$ &              $38$ &             $9.1$ &              $83$ &              $9.1$ &      $10$ & $9.9\times10^{-3}$ \\
J2142.4$+$3659 &                          - &             - &                - &   $4.1\pm0.9$ &     $2.6\pm1.0$ &              $16$ &              $40$ &              $19$ &             $0.53$ &     $1.1$ &             $0.76$ \\
J2246.7$-$5207 &                          - &             - &                - & $3.25\pm0.22$ &     $4.3\pm0.5$ &              $76$ & $1.4\times10^{2}$ &              $80$ &             $0.13$ &     $2.4$ &               $11$ \\
J2251.7$-$3208 &                          - &             - &                - &   $3.4\pm2.1$ & $0.3 [\dagger]$ & $2.3\times10^{2}$ & $4.2\times10^{2}$ & $2.4\times10^{2}$ &             $0.35$ &     $7.0$ &               $12$ \\
\bottomrule
\end{tabular}
\end{table}
\end{landscape}

\end{appendix}

\bsp	
\label{lastpage}
\end{document}